\documentclass[a4paper,11pt]{article}
\pdfoutput=1

\usepackage{amsmath,amssymb}
\usepackage[amsmath,thmmarks]{ntheorem}
\usepackage{color}
\usepackage[normalem]{ulem}
\usepackage{xspace}
\usepackage{booktabs, threeparttable, tablefootnote}
\usepackage{enumitem}
\usepackage{changepage}
\usepackage{authblk}

\usepackage{url,hyperref}
\usepackage[framemethod=TikZ]{mdframed}

\usepackage{pgfplots, tikz}
\usetikzlibrary{arrows,calc,positioning}
\pgfplotsset{compat=newest}
\usepgfplotslibrary{units}
\usepgfplotslibrary{external}
\usepgfplotslibrary{groupplots}
\usepgfplotslibrary{statistics}
\usetikzlibrary{pgfplots.external}
\usepackage{filemod}
\newcommand{\includetikz}[2]{\includegraphics{#2}}

\newcommand\PlotNetwork[3]{%
  \def\oPos{#1}%
  \def\CIJ{#2}%
  \def\lbl{#3}%
  \node[draw=none, fill=none] (ori) at \oPos {};%
  \foreach \i in {0,1,...,7}%
  {%
    \pgfmathsetmacro{\angle}{\i*360/8}%
    \node[minimum size=\nodeR] (\i) at ($(ori) + (\angle:\netR)$) {};%
  }%
  \foreach \i in {0,1,...,7}%
  {%
    \foreach \j in {0,1,...,7}%
    {%
      \pgfmathtruncatemacro\w{\CIJ[\i][\j]}%
      \ifnum\w>\th\draw[conn]  (\j) -- (\i);\fi%
    }%
  }%
  \node[draw=none, fill=none] (l) at ($(ori) + (135:\netR*1.4)$) {\bfseries \lbl};%
}

\tikzset{var/.style={draw, circle, fill=blue!20, minimum size=0.7cm}}

\newcommand{\yes}{\checkmark}
\newcommand{\no}{\ensuremath{\boldsymbol\times}}

\DeclareMathOperator{\tr}{tr}
\def\given{\,|\,}

\newcommand{\dd} {\mathrm{d}}
\newcommand{\tix} {\tilde{x}}
\newcommand{\tiX} {\tilde{X}}
\newcommand{\trans} {\mathrm{T}}

\newcommand{\be} {\begin{equation}}
\newcommand{\ee}{\end{equation}}

\newcommand{\phiGeometric}{\ensuremath{\Phi_G}\xspace}

\newcommand{\erdosrenyi}{Erd\H{o}s-R\'enyi\xspace}

\newcounter{boxCounter}

\newenvironment{titledBox}[1]
  {\tikzexternaldisable \refstepcounter{boxCounter} \begin{mdframed}[frametitle={Box \arabic{section}.\arabic{boxCounter}: #1}, skipabove=10pt, nobreak=true]}
  {\end{mdframed} \tikzexternalenable}
\numberwithin{boxCounter}{section}

\begin{document}

\title{Measuring Integrated Information: Comparison of\\Candidate Measures in Theory and Simulation}
\author[1,*]{Pedro A.M. Mediano}
\author[2]{Anil K. Seth}
\author[2]{Adam B. Barrett}

\affil[1]{Department of Computing, Imperial College, London, UK}
\affil[2]{Sackler Centre for Consciousness Science and Department of Informatics, University of Sussex, Brighton, UK}
\affil[ ]{ }
\affil[*]{Corresponding author: \href{mailto:pmediano@imperial.ac.uk}{\nolinkurl{pmediano@imperial.ac.uk}}}

\date{}

\maketitle

\section*{Abstract}

Integrated Information Theory (IIT) is a prominent theory of consciousness that
has at its centre measures that quantify the extent to which a system generates
more information than the sum of its parts. While several candidate measures of
integrated information (`$\Phi$') now exist, little is known about how they
compare, especially in terms of their behaviour on non-trivial network models.
In this article we provide clear and intuitive descriptions of six distinct
candidate measures. We then explore the properties of each of these measures in
simulation on networks consisting of eight interacting nodes, animated with
Gaussian linear autoregressive dynamics. We find a striking diversity in the
behaviour of these measures -- no two measures show consistent agreement across
all analyses. Further, only a subset of the measures appear to genuinely
reflect some form of dynamical complexity, in the sense of simultaneous
segregation and integration between system components. Our results help guide
the operationalisation of IIT and advance the development of measures of
integrated information that may have more general applicability.

\section{Introduction}

Since the seminal work of Tononi, Sporns and Edelman \cite{Tononi1994}, and
more recently, of Balduzzi and Tononi \cite{Balduzzi2008a}, there have been
many valuable contributions in neuroscience towards understanding and
quantifying the \textit{dynamical complexity} of a wide variety of systems. A
system is said to be dynamically complex if it shows a balance between two
competing tendencies, namely
\begin{itemize}\itemsep0pt
  \item \textbf{integration}, i.e.~the system behaves as one; and
  \item \textbf{segregation}, i.e.~the parts of the system behave independently.
\end{itemize}
The notion of dynamical complexity has also been variously described as a
balance between order and disorder, or between chaos and synchrony, and has
been related to criticality and metastability \cite{Mediano2016}. Many
quantitative measures of dynamical complexity have been proposed, but a
theoretically-principled, one-size-fits-all measure remains elusive.

A prominent framework highlighting the extent of simultaneous integration and
segregation is \textit{Integrated Information Theory} (IIT), which studies
dynamical complexity from information-theoretic principles. Measures of
integrated information attempt to quantify the extent to which the whole system
is generating more information than the `sum of its parts'. The information to
be quantified is typically the information that the current state contains
about a past state (for the information integrated over time window $\tau$, the
past state to be considered is that at time $\tau$ from the present). The
partitioning is done such that one considers the parts with the weakest links
between them, in other words, the partition across which integrated information
is computed is the `minimum information partition.' There are many ways one can
operationalise this concept of integrated information. Consequently, there now
exists a range of distinct integrated information measures.

Proponents of IIT claim that measures of integrated information potentially
relate to the quantity of consciousness generated by any physical system
\cite{Oizumi2014}. This is however controversial, and empirical evidence of a
relationship between any particular measure of integrated information and
consciousness remains scarce \cite{Cerullo2015}. Here, we do not focus on the
connections of IIT to consciousness, although we do comment on the application
of IIT to neural data (see Discussion). We instead consider measures of
integrated information more generally as useful operationalisations of notions
of dynamical complexity.

We have two goals. First, to provide a unified source of explanation of the
principles and practicalities of the various candidate measures of integrated
information. Second, to examine the behaviour of candidate measures on
non-trivial network models, in order to shed light on their comparative
practical utility.

In a recent related paper, Tegmark \cite{Tegmark2016} developed a theoretical
taxonomy of all integrated information measures that can be written as a
distance between a probability distribution pertaining to the whole and that
obtained from the product of probability distributions pertaining to the parts.
Here we review in detail five distinct and prominent proposed measures of
integrated information, including two ($\psi$ and $\Phi_G$) that were not
covered in Tegmark's taxonomy. These are: whole-minus-sum integrated
information $\Phi$ \cite{Balduzzi2008a}; integrated stochastic interaction
$\tilde{\Phi}$ \cite{Barrett2011}; integrated synergy $\psi$
\cite{Griffith2014}; decoder-based integrated information $\Phi^*$
\cite{Oizumi2015a}; geometric integrated information $\Phi_G$
\cite{Oizumi2015}. We also consider, for comparison, the measure causal density
(CD) \cite{Seth2011}, which can be considered as the sum of independent
information transfers in the system (without reference to a minimum information
partition). This measure has previously been discussed in conjunction with
integrated information measures \cite{Seth2006,Seth2011}.

All of the measures have the potential to behave in ways which are not obvious
\textit{a priori}, and in a manner difficult to express analytically. While
some simulations of some of the measures ($\Phi$, $\tilde{\Phi}$ and CD) on
networks have been performed \cite{Barrett2011,Seth2011}, other measures
($\Phi^*$ and $\Phi_G$) have not previously been computed on any model
consisting of more than two components. This paper provides a comparison of the
full suite of measures on non-trivial network models. We consider eight-node
networks with a range of different architectures, animated with basic noisy
vector autoregressive dynamics. We examine how network topology as well as
coupling strength and correlation of noise inputs affect each measure. We also
plot the relation between each measure and the global correlation (a simple
dynamical control). Based on these comparisons we discuss the extent to which
each measure appears genuinely to capture the co-existence of integration and
segregation central to the concepts of dynamical complexity and integrated
information.

After covering the necessary preliminaries in Section \ref{sec:preliminaries},
Section \ref{sec:integrationMeasures} sets out the intuition behind the
measures, and summarises the mathematics behind the definition of each measure.
In Section \ref{sec:results} we present the simulations. Then Section 5 is the
Discussion. In the Appendix, Section A.1, we derive new formulae for computing
the decoder-based integrated information $\Phi^*$ for Gaussian systems,
correcting the previous formulae in Ref.~\cite{Oizumi2015a}. Other Appendices
contain further derivations of mathematical properties of the measures.

\section{Notation, convention and preliminaries}
\label{sec:preliminaries}

In this section we review the fundamental concepts needed to define and discuss
the candidate measures of integrated information. In general, we will denote
random variables with uppercase letters (e.g. $X$, $Y$) and particular
instantiations with the corresponding lowercase letters (e.g. $x$, $y$).
Variables can be either continuous or discrete, and we assume that continuous
variables can take any value in $\mathbb{R}^n$ and that a discrete variable $X$
can take any value in the finite set $\Omega_X$. Whenever there is a sum
involving a discrete variable $X$ we assume the sum runs for all possible
values of $X$ (i.e. the whole $\Omega_X$). A partition $\mathcal{P} = \{M^1,
M^2, \dots, M^r\}$ divides the elements of system $X$ into $r$ non-overlapping,
non-empty sub-systems (or parts), such that $X = M^1 \bigcup M^2 \bigcup \dots
\bigcup M^r$ and $M^i
\bigcap M^j = \emptyset$, for any $i,j$. We denote each variable in $X$ as
$X^i$, and the total number of variables in $X$ as $n$. When dealing with time
series, time will be indexed with a subscript, e.g. $X_t$.

\textit{Entropy} $H$ quantifies the uncertainty associated with random variable
$X$ -- i.e. the higher $H(X)$ the harder it is to make predictions about $X$ --
and is defined as
\begin{align}
  H(X) =: - \displaystyle\sum_x p(x) \log p(x) ~ .
  \label{eq:ent}%
\end{align}
In many scenarios, a discrete set of states is insufficient to represent a
process or time series. This is the case, for example, with brain recordings,
which come in real-valued time series and with no \textit{a priori}
discretisation scheme. In these cases, using a continuous variable $X \in
\mathbb{R}$ we can similarly define the \textit{differential entropy},
\begin{align}
  H[p] =: -\int p(x) \log p(x) dx ~ .
  \label{eq:diffEnt}%
\end{align}
However, differential entropy is not as interpretable and well-behaved as its
discrete-variable counterpart. For example, differential entropy is not
invariant to rescaling or other transformations on $X$. Moreover, it is only
defined if $X$ has a density with respect to the Lebesgue measure $dx$; this
assumption will be upheld throughout this paper. We can also define the
\textit{conditional} and \textit{joint} entropies as
\begin{align}
\begin{split}
  H(X \given Y) & =: \sum_y p(y) H(X \given Y = y) \\
                & = -\sum_y p(y) \sum_x p(x \given y) \log p(x \given y)
  \label{eq:condEnt}%
\end{split}
\end{align}
\begin{align}
  H(X, Y) =: - \sum_{x,y} p(x, y) \log p(x, y) ~ ,
  \label{eq:jointEnt}%
\end{align}
\noindent respectively. Conditional and joint entropies can be analogously
defined for continuous variables by appropriately replacing sums with
integrals.

The Kullback-Leibler (KL) divergence quantifies the dissimilarity between two
probability distributions $p$ and $q$:
\begin{equation}
  D_{KL}(p\|q) =: \sum_x p(x) \log \frac{p(x)}{q(x)} ~ .
  \label{eq:kl}%
\end{equation}
The KL divergence represents a notion of (non-symmetric) distance between two
probability distributions. It plays an important role in information geometry,
which deals with the geometric structure of manifolds of probability
distributions.

Finally, mutual information $I$ quantifies the interdependence between two
random variables $X$ and $Y$. It is the KL divergence between the full joint
distribution and the product of marginals, but it can also be expressed as the
average reduction in uncertainty about $X$ when $Y$ is given:
\begin{align}
\begin{split}
  I(X; Y) & =: D_{KL} \left( p(X,Y) ~ \| ~ p(X) p(Y) \right) \\
          & = H(X) + H(Y) - H(X, Y) \\
          & = H(X) - H(X|Y) ~ .
  \label{eq:mi}%
\end{split}
\end{align}
Mutual information is symmetric in the two arguments $X$ and $Y$. We make use
of the following properties of mutual information:
\begin{enumerate}\itemsep0pt
  \item $I(X; Y) = I(Y; X)$,
  \item $I(X; Y) \geq 0$, and
  \item $I(f(X); g(Y)) = I(X; Y)$ for any injective functions $f, g$.
\end{enumerate}
We highlight one implication of property 3: $I$ is upper-bounded by the entropy
of both $X$ and $Y$. This means that the entropy $H(X)$ of a random variable
$X$ is the maximum amount of information $X$ can have about any other variable
$Y$ (or another variable $Y$ can have about $X$).

Mutual information is defined analogously for continuous variables and, unlike
differential entropy, it retains its interpretability in the continuous
case.\footnote{The formal derivation of the differential entropy
  proceeds by considering the entropy of a discrete variable with $k$ states, and taking
  the $k \rightarrow \infty$ limit. The result is the differential entropy plus a divergent
  term that is usually dropped and is ultimately responsible for the undesirable properties
  of differential entropy. In the case of $I(X;Y)$ the divergent terms for the various
entropies involved cancel out, restoring the useful properties of its discrete
counterpart \cite{Cover2006}.} Furthermore, one can track how much information
a system preserves during its temporal evolution by computing the time-delayed
mutual information (TDMI) $I(X_t; X_{t-\tau})$.

Next, we introduce notation and several useful identities to handle Gaussian
variables. Given an $n$-dimensional real-valued system $X$, we denote its
covariance matrix as $\Sigma(X)_{ij} =: \mathrm{cov}(X^i, X^j)$. Similarly,
cross-covariance matrices are denoted as $\Sigma(X,Y)_{ij} =: \mathrm{cov}(X^i,
Y^j)$. We will make use of the conditional (or partial) covariance formula,
\begin{align}
  \Sigma(X|Y) =: \Sigma(X) - \Sigma(X,Y) \Sigma(Y)^{-1} \Sigma(Y, X) ~ .
  \label{eq:partialCov}%
\end{align}
\noindent For Gaussian variables,
\begin{align}
  H(X) & = \frac{1}{2} \log ( \det \Sigma(X) ) + \frac{1}{2} n \log (2\pi e)\,, \label{eq:Gaussentropy} \\
  H(X|Y=y) & = \frac{1}{2}\log ( \det\Sigma(X|Y)  ) + \frac{1}{2} n \log (2\pi e)\,,\,\,\, \forall y\,,\label{eq:Gausscondent}\\
  I(X;Y) & = \frac{1}{2}\log \left(\frac{ \det\Sigma(X) }{\det\Sigma(X|Y) }\right)\,. \label{eq:Gaussinf}
\end{align}
All systems we deal with in this article are stationary and ergodic, so
throughout the paper $\Sigma(X_t) =\Sigma(X_{t-\tau})$ for any $\tau$.

\section{Integrated information measures}
\label{sec:integrationMeasures}

\subsection{Overview}
\label{sec:overview}

In this section we review the theoretical underpinnings and practical
considerations of several proposed measures of integrated information, and in
particular how they relate to intuitions about segregation, integration and
complexity. These measures are:

\begin{itemize}\itemsep0pt
  \item Whole-minus-sum integrated information, $\Phi$;
  \item Integrated stochastic interaction, $\tilde{\Phi}$;
  \item Integrated synergy, $\psi$;
  \item Decoder-based integrated information, $\Phi^*$;
  \item Geometric integrated information, $\Phi_G$; and
  \item Causal density, CD.
\end{itemize}

All of these measures (besides CD) have been inspired by the measure proposed
by Balduzzi and Tononi in \cite{Balduzzi2008a}, which we call
$\Phi_{2008}$\footnote{Causal density is somewhat distinct, but is still a
measure of complexity based on information dynamics between the past and
current state; therefore its inclusion here will be useful.}. $\Phi_{2008}$ was
based on the information the current state contains about a hypothetical
maximum entropy past state.
In practice, this results in measures that are applicable only to discrete
Markovian systems \cite{Barrett2011}. For broader applicability, it is more
practical to build measures based on the ongoing spontaneous information
dynamics -- that is, based on $p(X_t,X_{t-\tau})$ without applying a
perturbation to the system. Measures are then well-defined for any stochastic
system (with a well-defined Lebesgue measure across the states), and can be
estimated for real data using empirical distributions if stationarity can be
assumed. All of the measures we consider in this paper are based on a system's
spontaneous information dynamics.

Table \ref{tab:measureReferences} contains a brief description of each measure
and a reference to the original publication that introduced it. We refer the
reader to the original publications for more detailed descriptions of each
measure. Table \ref{tab:overview} contains a summary of properties of the
measures considered, proven for the case in which the system is ergodic and
stationary, and the spontaneous distribution is used.

\begin{table}[!ht]
\centering
\caption{Integrated information measures considered and original references.}
\label{tab:measureReferences}
\begin{tabular}{c|l|c}
  \textbf{Measure} & \textbf{Description} & \textbf{Reference} \\
  \toprule
  $\Phi$        & Information lost after splitting the system & \cite{Balduzzi2008a} \\
  $\tilde{\Phi}$     & Uncertainty gained after splitting the system & \cite{Barrett2011} \\
  $\psi$        & Synergistic predictive information between parts of the system & \cite{Griffith2014} \\
  $\Phi^*$      & Past state decoding accuracy lost after splitting the system & \cite{Oizumi2015a} \\
  $\Phi_G$ & Information-geometric distance to system with disconnected parts & \cite{Oizumi2015} \\
  CD            & Average pairwise directed information flow & \cite{Seth2011}\tablefootnote{Although the origins of causal density go as back as \cite{Granger1969}, it hasn't been until the last decade that it has found its way into neuroscience. The paper referenced in the table acts as a modern review of the properties and behaviour of causal density.} \\
\end{tabular}
\end{table}

\begin{table}[ht]
\centering
\caption{Overview of properties of integrated information measures. Proofs in Appendix \ref{ap:proofs}.}
\label{tab:overview}
\begin{tabular}{l|c|c|c|c|c|c}
  ~ & $\Phi$ & $\tilde{\Phi}$ & $\psi$ & $\Phi^*$ & $\Phi_G$ & CD \\
  \toprule
  Time-symmetric & \yes & \yes & \no & \no & \yes & \no \\
  Non-negative & \no & \yes & \yes & \yes & \yes & \yes \\
  Invariant to variable rescaling & \yes & \no & \yes & \yes & \yes & \yes \\
  Upper-bounded by time-delayed mutual information & \yes & \no & \yes & \yes & \yes & \yes \\
  Computable for arbitrary real-valued systems & \yes & \yes & \no & \no & \no & \yes \\
  Closed-form expression in discrete and Gaussian systems & \yes & \yes & \yes & \no & \no & \yes \\
\end{tabular}
\end{table}

\subsection{Minimum information partition}
\label{sec:mip}

Key to all measures of integrated information is the notion of splitting or
partitioning the system to quantify the effect of such split on the system as a
whole. In that spirit, integrated information measures are defined through some
measure of \textit{effective information}, which operationalises the concept of
``information \textit{beyond} a partition'' $\mathcal{P}$. This typically
involves splitting the system according to $\mathcal{P}$ and computing some
form of information loss, via (for example) mutual information ($\Phi$),
conditional entropy ($\tilde{\Phi}$), or decoding accuracy ($\Phi^*$) (see
Table \ref{tab:measureReferences}). Integrated information is then the
effective information with respect to the partition that identifies the
``weakest link'' in the system, i.e. the partition for which the parts are
least integrated. Formally, integrated information is the effective information
beyond the \textit{minimum information partition} (MIP), which, given an
effective information measure $f[X; \tau, \mathcal{P}]$, is defined as
\begin{align}
  \mathcal{P}_{\mathrm{MIP}} = \arg_{\mathcal{P}}\min \frac{f[X; \tau, \mathcal{P}]}{K(\mathcal{P})} ~ ,
\end{align}
\noindent where $K(\mathcal{P})$ is a normalisation coefficient. In other
words, the MIP is the partition across which the (normalised) effective
information is minimum, and integrated information is the (unnormalised)
effective information beyond the MIP. The purpose of the normalisation
coefficient is to avoid biasing the minimisation towards unbalanced
bipartitions (recall that the extent of information sharing between parts is
bounded by the entropy of the smaller part). Balduzzi and Tononi
\cite{Balduzzi2008a} suggest the form
\begin{align}
  K(\mathcal{P}) = (r-1) \min_k H(M_t^k) ~ .
\end{align}
However, not all contributions to IIT have followed Balduzzi and Tononi's
treatment of the MIP. Of the measures listed above, $\Phi$ and $\tilde{\Phi}$
share this partition scheme, $\psi$ defines the MIP through an
\textit{unnormalised} effective information, and $\Phi^*$, $\Phi_G$ and CD are
defined via the atomic partition without any reference to the MIP. These
differences are a confounding factor when it comes to comparing measures -- it
becomes difficult to ascertain whether differences in behaviour of various
measures are due to their definitions of effective information, to their
normalisation factor (or lack thereof), or to their partition schemes. We
return to this discussion in Sec.~\ref{sec:partitions}.

In the following we present all measures as they were introduced in their
original papers (see Table \ref{tab:measureReferences}), although it is trivial
to combine different effective information measures with different partition
optimisation schemes. However, all results presented in
Sec.~\ref{sec:results} are calculated by minimising each unnormalised effective
information measure over even-sized bipartitions -- i.e. bipartitions in which
both parts have the same number of components. This is to avoid conflating the
effect of the partition scan method with the effect of the integrated
information measure itself.

\subsection{Whole-minus-sum integrated information $\Phi$}
\label{sec:phi2008}

We next turn to the different measures of integrated information. As
highlighted above, a primary difference among them is how they define the
effective information beyond a given partition. Since most measures were
inspired by Balduzzi and Tononi's $\Phi_{2008}$, we start there.

For $\Phi_{2008}$, the effective information $\varphi_{2008}$ is given by the
KL divergence between $p_{\mathrm{c}}(X_0|X_1=x)$ and $\Pi_k
p_{\mathrm{c}}(M_0^k|M_1^k=m^k)$, where $p_c(X_0|X_1=x)$ (and analogously
$p_{\mathrm{c}}(M_0^k|M_1^k=m^k)$) is the conditional distribution for $X_0$
given $X_1=x$ under the perturbation at time 0 into all states with equal
probability -- i.e.~given that the joint distribution is given by
$p_{\mathrm{ce}}(X_0,X_1)=:p(X_1|X_0)p_u(X_0)$, where $p_u$ is the uniform
(maximum entropy) distribution\footnote{Here we follow notation from
  \cite{Krohn2016}. The c and e here stand respectively for cause and effect.
  Without an initial condition, here that the uniform distribution holds at
  time 0, there would be no well-defined probability distribution for these
  states.  Further, Markovian dynamics are required for these probability
  distributions to be well-defined; for non-Markovian dynamics, a longer chain
of initial states would have to be specified, going beyond just that at time
0.}.

Averaging $\varphi_{2008}$ over all states $x$, the result can be expressed as
either
\begin{align}
  I(X_0; X_1) - \sum_{k=1}^r I(M_0^k; M_1^k) ~ ,
\end{align}
\noindent or
\begin{align}
  -H(X_0 | X_1) + \sum_{k=1}^r H(M_0^k | M_1^k) ~ .
\end{align}
These two expressions are equivalent under the uniform perturbation, since they
differ only by a factor that vanishes if $p(X_0)$ is the uniform distribution.
However, they are \textit{not} equivalent if the spontaneous distribution of
the system is used instead -- i.e. if $p(X_{t-\tau}, X_t)$ is used instead of
$p_{\mathrm{ce}}(X_0, X_1)$. This means that for application to spontaneous
dynamics (i.e. without perturbation) we have two alternatives that give rise to
two measures that are both equally valid analogs of $\Phi_{2008}$.

We call the first alternative whole-minus-sum integrated information $\Phi$
($\Phi_{\mathrm{E}}$ in \cite{Barrett2011}). The effective information
$\varphi$ is defined as the difference in time-delayed mutual information
between the whole system and the parts. The effective information of the system
beyond a certain partition $\mathcal{P}$ is
\begin{align}
  \varphi[X; \tau, \mathcal{P}] =: I(X_{t-\tau}; X_t) - \displaystyle\sum_{k=1}^r I(M_{t-\tau}^k; M_t^k) ~ .
  \label{eq:effectiveInfo}%
\end{align}
We can interpret $I(X_t; X_{t-\tau})$ as how good the system is at predicting
its own future or decoding its own past\footnote{Future and past are equivalent
because mutual information is symmetric.}. Then $\varphi$ here can be seen as
the loss in predictive power incurred by splitting the system according to
$\mathcal{P}$. The details of the calculation of $\Phi$ (and the MIP) are shown
in Box \ref{box:phi}.

$\Phi$ is often regarded as a poor measure of integrated information because it
can be negative \cite{Oizumi2015a}. This is indeed conceptually awkward if
$\Phi$ is seen as an absolute measure of integration between the parts of a
system, though it is a reasonable property if $\Phi$ is interpreted as a ``net
synergy'' measure \cite{Barrett2014} -- quantifying to what extent the parts
have shared or complementary information about the future state. That is, if
$\Phi > 0$ we infer that the whole is better than the parts at predicting the
future (i.e., $\Phi > 0$ is a sufficient condition), but a negative or zero
$\Phi$ does not imply the opposite. Therefore, from an IIT perspective a
negative $\Phi$ can lead to the understandably confusing interpretation of a
system having ``negative integration,'' but through a different lens (net
synergy) it can be more easily interpreted as (negative) overall redundancy in
the evolution of the system. See Section \ref{sec:psi} and
Ref.~\cite{Barrett2014} for further discussion on whole-minus-sum measures.

\begin{titledBox}{Calculating whole-minus-sum integrated information $\Phi$}
  \begin{subequations}
  \begin{gather}
    \Phi[X;\tau] = \varphi[X; \tau, \mathcal{B}^{\mathrm{MIB}}] \\
    \mathcal{B}^{\mathrm{MIB}} = \arg_{\mathcal{B}}\min \frac{\varphi[X; \tau, \mathcal{B}]}{K(\mathcal{B})} \label{eq:phiMIB} \\
    \varphi[X; \tau, \mathcal{B}] = I(X_{t-\tau}; X_t) - \displaystyle\sum_{k=1}^2 I(M_{t-\tau}^k; M_t^k) \\
    K(\mathcal{B}) = \min \left\{ H(M_t^1), H(M_t^2) \right\}
  \end{gather}
  \label{eq:phi}
  \end{subequations}
  \vspace{-20pt}
  \begin{enumerate}
    \item For \textbf{discrete variables}:
    \begin{align*}
      I(X_{t-\tau}; X_t) = \displaystyle\sum_{x, x'} p(X_{t-\tau} = x, X_t = x') \log \left( \frac{p(X_{t-\tau} = x, X_t = x')}{p(X_{t-\tau} = x) ~ p(X_t = x')} \right)
    \end{align*}
    \item For \textbf{continuous, linear-Gaussian variables}:
    \begin{align*}
      I(X_{t-\tau}; X_t) = \frac{1}{2} \log \left(\frac{\det \Sigma(X_t)}{\det \Sigma(X_t \given X_{t-\tau})} \right)
    \end{align*}
    \item For \textbf{continuous variables} with an arbitrary distribution, we
    must resort to the nearest-neighbour methods introduced by
    \cite{Kraskov2004}. See reference for details.
  \end{enumerate}
  \label{box:phi}
\end{titledBox}

\subsection{Integrated stochastic interaction $\tilde{\Phi}$}
\label{sec:phiTilde}

We next consider the second alternative for $\Phi_{2008}$ for spontaneous
information dynamics: integrated stochastic interaction $\tilde{\Phi}$. Also
introduced in Barrett and Seth \cite{Barrett2011}, this measure embodies
similar concepts as $\Phi$, with the main difference being that $\tilde{\Phi}$
utilises a definition of effective information in terms of an \textit{increase
in uncertainty} instead of in terms of a \textit{loss of information}.

$\tilde{\Phi}$ is based on \textit{stochastic interaction} $\tilde\varphi$,
introduced by Ay \cite{Ay2015}. Akin to Eq.~\eqref{eq:effectiveInfo}, we define
stochastic interaction beyond partition $\mathcal{P}$ as
\begin{align}
  \tilde\varphi[X; \tau, \mathcal{P}] =: \displaystyle\sum_{k=1}^r H(M_{t-\tau}^k \given M_t^k) - H(X_{t-\tau} \given X_t) ~ .
  \label{eq:stochasticInteraction}%
\end{align}
Stochastic interaction quantifies to what extent uncertainty about the past is
increased when the system is split in parts, compared to considering the system
as a whole. The details of the calculation of $\tilde{\Phi}$ are similar to
those of $\Phi$ and are described in Box \ref{box:phiTilde}.

The most notable advantage of $\tilde{\Phi}$ over $\Phi$ as a measure of
integrated information is that $\tilde{\Phi}$ is guaranteed to be non-negative.
In fact, as mentioned above $\varphi$ and $\tilde\varphi$ are related through
the equation
\begin{align}
  \tilde\varphi[X; \tau, \mathcal{P}] = \varphi[X; \tau, \mathcal{P}] + I(M_t^1; M_t^2; \ldots; M_t^r) ~ ,
\end{align}
\noindent where
\begin{align}
  I(M_t^1; M_t^2; \ldots; M_t^r) =  \sum_{k=1}^r H(M_t^k) - H(X_t) ~ .
\end{align}
This measure is also linked to \textit{information destruction}, as presented
in Wiesner et al. \cite{Wiesner2011}. The quantity $H(X_{t-\tau} | X_t)$
measures the amount of irreversibly destroyed information, since $H(X_{t-\tau}
| X_t) > 0$ indicates that more than one possible past trajectory of the system
converged on the same present state, making the system irreversible and
indicating a loss of information about the past states. From this perspective,
$\tilde\varphi$ can be understood as the difference between the information
that is considered destroyed when the system is observed as a whole, or split
into parts. Note however that this measure is time-symmetric when applied to a
stationary system; for stationary systems total instantaneous entropy does not
increase with time.

\begin{titledBox}{Calculating integrated stochastic interaction $\tilde{\Phi}$}
  \begin{subequations}
  \begin{gather}
    \tilde\Phi[X;\tau] = \tilde\varphi[X; \tau, \mathcal{B}^{\mathrm{MIB}}] \\
    \mathcal{B}^{\mathrm{MIB}} = \arg_{\mathcal{B}}\min \frac{\tilde\varphi[X; \tau, \mathcal{B}]}{K(\mathcal{B})} \\
    \tilde\varphi[X; \tau, \mathcal{B}] = \displaystyle\sum_{k=1}^r H(M_{t-\tau}^k \given M_t^k) - H(X_{t-\tau} \given X_t) \\
    K(\mathcal{B}) = \min \left\{ H(M_t^1), H(M_t^2) \right\}
  \end{gather}
  \label{eq:phiTilde}
  \end{subequations}
  \vspace{-20pt}
  \begin{enumerate}
    \item For \textbf{discrete variables}:
    \begin{align*}
      H(X_{t-\tau} \given X_t) = - \displaystyle\sum_{x, x'} p(X_{t-\tau} = x, X_t = x') \log \left( \frac{p(X_{t-\tau} = x, X_t = x')}{p(X_t = x')} \right)
    \end{align*}
    \item For \textbf{continuous, linear-Gaussian variables}:
    \begin{align*}
      H(X_{t-\tau} \given X_t) = \frac{1}{2} \log \det \Sigma(X_{t-\tau} \given X_t) + \frac{1}{2}n\log(2\pi e)
    \end{align*}
    \item For \textbf{continuous variables} with an arbitrary distribution, we
    must resort to the nearest-neighbour methods introduced by
    \cite{Kraskov2004}. See reference for details.
  \end{enumerate}
  \label{box:phiTilde}
\end{titledBox}

\subsection{Integrated synergy $\psi$}
\label{sec:psi}

Originally designed as a ``more principled'' integrated information measure
\cite{Griffith2014}, $\psi$ shares some features with $\Phi$ and $\tilde{\Phi}$
but is grounded in a different branch of information theory, namely the Partial
Information Decomposition (PID) framework, as described by Williams and Beer
\cite{Williams2010}. In the PID, the information that two (source) variables
provide about a third (target) variable is decomposed into four non-negative
terms as
\begin{align*}
  I(X,Y ; Z) = U_X(X; Z) + U_Y(Y; Z) + R({X,Y}; Z) + S({X, Y}; Z) ~ ,
\end{align*}
\noindent where $U_\alpha$ is the \textit{unique information} of source $\alpha$,
$R$ is the \textit{redundancy} between both sources and $S$ is their
\textit{synergy}. Figure \ref{fig:pidVenn} illustrates the involved quantities in
a Venn diagram.

{ \tikzexternaldisable
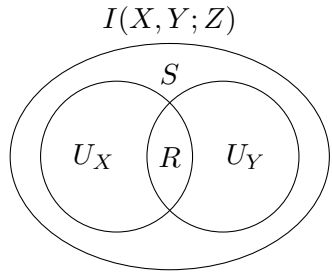
\begin{figure}[ht]
  \centering
  \begin{tikzpicture}
    \draw (-0.7cm,0) circle (1cm);
    \draw (0.7cm,0) circle (1cm);
    \draw (0,0) ellipse (2.1cm and 1.5cm);

    \draw (-1cm,0) node {$U_X$};
    \draw (1cm,0) node {$U_Y$};
    \draw (0,0) node {$R$};
    \draw (0, 1.1cm) node {$S$};
    \draw (0, 1.8cm) node {$I(X,Y; Z)$};
  \end{tikzpicture}
  \caption{Venn diagram of the partial information decomposition \cite{Williams2010}.}
  \label{fig:pidVenn}
\end{figure}
\tikzexternalenable
}

Integrated synergy $\psi$ is the information that the parts provide about the
future of the system that is exclusively synergistic -- i.e.~cannot be provided
by any combination of parts independently:
\begin{align}
  \psi[X; \tau, \mathcal{P}] =: I(X_{t-\tau}; X_t) - \max_\mathcal{P} I_\cup (M^1_{t-\tau}, M^2_{t-\tau}, \dots, M^r_{t-\tau} ; X_t) ~ ,
\end{align}
\noindent where
\begin{align}
  I_\cup (M^1_{t-\tau}, \ldots, M^r_{t-\tau} ; X_t) =: ~ ~ ~ \sum_{\makebox[0pt]{$\scriptstyle \mathcal{S}\subseteq \{ M^1, \ldots, M^r\}$}} ~ ~ ~ (-1)^{|\mathcal{S}|+1} I_\cap (\mathcal{S}^1_{t-\tau},\ldots,\mathcal{S}^{|\mathcal{S}|}_{t-\tau};X_t) ~ ,
\end{align}
\noindent and $I_\cap(\mathcal{S}_1,\ldots,\mathcal{S}_{|\mathcal{S}|};Z)$
denotes the redundant information sources
$\mathcal{S}_1,\ldots,\mathcal{S}_{|\mathcal{S}|}$ have about target $Z$.
The main problem of PID is that it is underdetermined. For example, for the
case of two sources, Shannon's information theory specifies three quantities
($I(X,Y;Z)$, $I(X;Z)$, $I(Y;Z)$) whereas PID specifies four ($S$, $R$, $U_X$,
$U_Y$). Therefore, a complete operational definition of $\psi$ requires a
definition of redundancy from which to construct the partial information
components \cite{Williams2010}. In this sense, the main shortcoming of $\psi$,
inherited from PID, is that there is no agreed consensus on a definition of
redundancy \cite{Barrett2014,Bertschinger2012}.

Here, we take Griffith's conceptual definition of $\psi$ and we complement it
with available definitions of redundancy. For the linear-Gaussian systems we
will be studying in Sec.~\ref{sec:results}, we use the minimum mutual
information PID presented in \cite{Barrett2014}\footnote{Barrett's derivation
of the MMI-PID, which follows Williams and Beer and Griffith and Koch's
procedure, gives this formula when the target is univariate. We generalise the
formula here to the case of multivariate target in order to render $\psi$
computable for Gaussians. This formula leads to synergy being the extra
information contributed by the weaker source given the stronger source was
previously known.}. Although we do not show any discrete examples here, for
completeness we provide complete formulae to calculate $\psi$ for discrete
variables using Griffith and Koch's redundancy measure \cite{Griffith2012}.
Note that alternatives are available for both discrete and linear-Gaussian
systems \cite{Rosas2016,Ince2017,Williams2010,Bertschinger2014,Kay2018}.

\begin{titledBox}{Calculating integrated synergy $\psi$}
  \begin{align}
    \psi[X; \tau, \mathcal{P}] = I(X_{t-\tau}; X_t) - \max_{\mathcal{P}} I_\cup (M^1_{t-\tau}, \ldots, M^r_{t-\tau} ; X_t)
    \label{eq:psi}
  \end{align}
  \vspace{-20pt}
  \begin{enumerate}
    \item For \textbf{discrete variables}: (following Griffith and Koch's \cite{Griffith2012} PID scheme)
    \begin{subequations}%
    \begin{alignat*}{2}%
      I_\cup(M^1_{t-\tau},\ldots,M^r_{t-\tau}; X_t) = & \displaystyle\min_q \displaystyle\sum_{x, x'} q(x, x') \log \left( \frac{q(x, x')}{q(x) ~ q(x')} \right) \\
      \mathrm{s.t.} ~ ~ & q(M^i_{t-\tau}, X_t) = p(M^i_{t-\tau}, X_t)
    \end{alignat*}%
    \end{subequations}%

    \item For \textbf{continuous, linear-Gaussian variables}:
    \begin{align*}
      I_\cup(M^1_{t-\tau},\ldots,M^r_{t-\tau} ; X_t) = \max_k I(M^k_{t-\tau}; X_t)
    \end{align*}
    
    \item For \textbf{continuous variables} with an arbitrary distribution: unknown.
  \end{enumerate}
  \label{box:psi}
\end{titledBox}

\subsection{Decoder-based integrated information $\Phi^*$}
\label{sec:phiStar}

Introduced by Oizumi et al. in Ref.~\cite{Oizumi2015a}, decoder-based
integrated information $\Phi^*$ takes a different approach from the previous
measures. In general, $\Phi^*$ is given by
\begin{align}
  \Phi^*[X; \tau, \mathcal{P}] =: I(X_{t-\tau}; X_t) - I^*[X; \tau, \mathcal{P}] ~ ,
\end{align}
\noindent where $I^*$ is known as the \textit{mismatched decoding information},
and quantifies how much information can be extracted from a variable if the
receiver is using a suboptimal (or \textit{mismatched}) decoding distribution
\cite{Latham2005,Merhav1994}. This mismatched information has been used in
neuroscience to quantify the contribution of neural correlations in stimulus
coding \cite{Oizumi2010}, and can similarly be used to measure the contribution
of inter-partition correlations to predictive information.

To calculate $\Phi^*$ we formulate a restricted model $q$ in which the
correlations between partitions are ignored,
\begin{align}
  q(X_t | X_{t - \tau}) = \displaystyle \prod_i p(M_t^i | M_{t-\tau}^i) ~ ,
\end{align}
\noindent and we calculate $I^*$ for the case where the sender is using the
full model $p$ as an encoder and the receiver is using the restricted model $q$
as a decoder. The details of the calculation of $\Phi^*$ and $I^*$ are shown in
Box \ref{box:phiStar}. Unlike the previous measures shown in this section,
$\Phi^*$ does not have an interpretable formulation in terms of simpler
information-theoretic functionals like entropy and mutual information.

Calculating $I^*$ involves a one-dimensional optimisation problem, which is
straightforwardly solvable if the optimised quantity, $\tilde{I}(\beta)$, has a
closed form expression \cite{Latham2005}. For systems with continuous
variables, it is in general very hard to estimate $\tilde{I}(\beta)$. However,
for continuous linear-Gaussian systems and for discrete systems
$\tilde{I}(\beta)$ has an analytic closed form as a function of $\beta$ if the
covariance or joint probability table of the system are known, respectively. In
Appendix \ref{ap:phiStar} we derive the formulae. (Note the version written
down in \cite{Oizumi2015a} is incorrect, although their simulations match our
results; we checked results from our derived version of the formulae versus
results obtained from numerical integration, and confirmed that our derived
formulae are the correct ones.)
Conveniently, in both the discrete and the linear-Gaussian case
$\tilde{I}(\beta)$ is concave in $\beta$ (proofs in \cite{Latham2005} and in
Appendix \ref{ap:phiStar}, respectively), which makes the optimisation
significantly easier.

\begin{titledBox}{Calculating decoder-based integrated information $\Phi^*$}
  \begin{subequations}
    \begin{gather}
      \Phi^*[X; \tau, \mathcal{P}] = I(X_{t-\tau}; X_t) - I^*[X; \tau, \mathcal{P}] \\
      I^*[X; \tau, \mathcal{P}] = \max_\beta \tilde{I}(\beta; X, \tau, \mathcal{P})
    \end{gather}
    \label{eq:phiStar}
  \end{subequations}
  \vspace{-20pt}
  \begin{enumerate}
    \item For \textbf{discrete variables}:
    \begin{multline*}
      \tilde{I}(\beta; X, \tau, \mathcal{P}) = - \displaystyle\sum_{x'} p(X_t = x')
      \log \displaystyle\sum_{x} p(X_{t-\tau} = x) q(X_t = x' | X_{t-\tau} = x)^{\beta} \\ +
      \displaystyle\sum_{x, x'} p(X_{t-\tau} = x, X_t = x') \log q(X_t = x' | X_{t-\tau} = x)^{\beta}
    \end{multline*}

    \item For \textbf{continuous, linear-Gaussian variables}: (see appendix for details)
    \begin{align*}
      \tilde{I}(\beta; X, \tau, \mathcal{P}) = \frac{1}{2} \log \left( |Q| |\Sigma_x| \right) +
      \frac{1}{2} \tr \left( \Sigma_x R \right) + \beta \tr \left( \Pi^{-1}_{x|\tilde{x}} \Pi_{x\tilde{x}} \Pi^{-1}_{x} \Sigma_{\tilde{x}x} \right)
    \end{align*}
    
    \item For \textbf{continuous variables} with an arbitrary distribution: unknown.
  \end{enumerate}
  \label{box:phiStar}
\end{titledBox}

\subsection{Geometric integrated information $\Phi_G$}

In \cite{Oizumi2015}, Oizumi et al. approach the notion of dynamical complexity
via yet another formalism. Their approach is based on \textit{information
geometry} \cite{Amari2000,Amari2010}. The objects of study in information
geometry are spaces of families of probability distributions, considered as
differentiable (smooth) manifolds. The natural metric in information geometry
is the Fisher information metric, and the KL divergence provides a natural
measure of (asymmetric) distance between probability distributions. Information
geometry is the application of differential geometry to the relationships and
structure of probability distributions.

To quantify integrated information, Oizumi et al. \cite{Oizumi2015} consider
the divergence between the complete model of the system under study
$p(X_{t-\tau}, X_t)$ and a \textit{restricted model} $q(X_{t-\tau}, X_t)$ in
which links between the parts of the system have been severed. This is known as
the \textit{M-projection} of the system onto the manifold of restricted models
$Q = \{q ~ \colon q(M^i_t | X_{t-\tau}) = q(M^i_t | M^i_{t-\tau}) \} $, and
\begin{align}
  \Phi_G[X; \tau, \mathcal{P}] =: \displaystyle\min_{q \in Q} D_{KL} \left(p(X_{t-\tau},X_t) \| q(X_{t-\tau},X_t)\right) ~ .
\end{align}
Key to this measure is that in considering the partitioned system, it is only
the connections that are cut; correlations between the parts are still allowed
on the partitioned system. Although conceptually simple, $\Phi_G$ is very hard
to calculate compared to all other measures we consider here (see Box
\ref{box:phiG}). There is no known closed form solution for any system, and we
can only find approximate numerical estimates for some systems. In particular,
for discrete and linear-Gaussian variables we can formulate $\Phi_G$ as the
solution of a pure constrained multivariate optimisation problem, with the
advantage that the optimisation objective is differentiable and convex
\cite{Boyd2004}.

\begin{titledBox}{Calculating geometric integration $\Phi_G$}
  \begin{subequations}%
  \begin{alignat}{2}%
    \Phi_G[X; \tau, \mathcal{P}] = & \displaystyle\min_q D_{KL} (p \| q) \\
    \mathrm{s.t.} ~ ~ & q(M^i_{t+\tau} \given X_t) = q(M^i_{t+\tau} \given M^i_t)
    \label{eq:phiG}%
  \end{alignat}%
  \end{subequations}%
  \vspace{-20pt}
  \begin{enumerate}
    \item For \textbf{discrete variables}: numerically optimise the objective $D_{KL} (p \| q)$
    subject to the constraints
    \begin{align*}
      \sum_{x,x'} q(X_{t-\tau} = x', X_t = x) = 1 \hspace{4ex} \mathrm{and} \hspace{4ex} q(M^i_t \given X_{t-\tau}) = q(M^i_t \given M^i_{t-\tau}) ~ \forall i
    \end{align*}

    \item For \textbf{continuous, linear-Gaussian variables}: numerically optimise the objective
    \begin{align*}
      \Phi_G[X; \tau, \mathcal{P}] = \min_{\Sigma(E)'} \frac{1}{2} \log \frac{|\Sigma(E)'|}{|\Sigma(E)|} ~ ,
    \end{align*}
    where $\Sigma(E) = \Sigma(X_t|X_{t-1})$, and subject to the constraints
    \begin{align*}
      & \Sigma(E)' = \Sigma(E) + (A-A')\Sigma(X)(A-A')^{\mathrm{T}} \hspace{5ex} \mathrm{and} \\
      & (\Sigma(X)(A-A')\Sigma(E)'^{-1})_{ii} = 0
    \end{align*}
    
    \item For \textbf{continuous variables} with an arbitrary distribution: unknown.
  \end{enumerate}
  \label{box:phiG}
\end{titledBox}

\subsection{Causal density}

Causal density (CD) is somewhat distinct from the other measures considered so far, in the sense that it is a sum of information transfers
rather than a direct measure of the extent to which the whole is greater than
the parts. Nevertheless, we include it here because of its relevance and use in the dynamical complexity literature.

CD was originally defined in terms of Granger causality \cite{Granger1969}, but
here we write it in terms of Transfer Entropy (TE) which provides a more
general information-theoretic definition \cite{Barnett2009}. The conditional
transfer entropy from $X$ to $Y$ conditioned on $Z$ is defined as
\begin{align}
  \mathrm{TE}_\tau(X \rightarrow Y \given Z) =: I(X_t; Y_{t+\tau} \given Z_t, Y_t) ~ .
  \label{eq:transferEnt}%
\end{align}
With this definition of TE we define CD as the average pairwise conditioned TE
between all variables in $X$,
\begin{align}
  \mathrm{CD}[X; \tau, \mathcal{P}] =: \frac{1}{r(r-1)} \displaystyle\sum_{i\neq j} \mathrm{TE}_\tau(M^i \rightarrow M^j \given M^{[ij]}) ,
\end{align}
\noindent where $M^{[ij]}$ is the subsystem formed by all variables in $X$
except for those in parts $M^i$ and $M^j$.

In a practical sense, CD has many advantages. It has been thoroughly studied in
theory \cite{Barnett2011} and applied in practice, with application
domains ranging from complex systems to neuroscience
\cite{Lindner2011,Lizier2010a,Mediano2017a}. Furthermore, there are
off-the-shelf algorithms that calculate TE in discrete and continuous systems
\cite{Barnett2014}. For details of the calculation of CD see Box \ref{box:cd}.

Causal density is a principled measure of dynamical complexity, as it vanishes
for purely segregated or purely integrated systems. In a highly segregated
system there is no information transfer at all, and in a highly integrated
system there is no transfer from one variable to another beyond the rest of the
system \cite{Seth2011}. Furthermore, CD is non-negative and upper-bounded by the
total time-delayed mutual information (proof in Appendix \ref{ap:cd}),
therefore satisfying what other authors consider an essential requirement for a
measure of integrated information \cite{Oizumi2015}.

\begin{titledBox}{Calculating causal density CD}
  \begin{align}
    \mathrm{CD}[X; \tau, \mathcal{P}] = \frac{1}{r(r-1)} \displaystyle\sum_{i \neq j} \mathrm{TE}_\tau(M^i \rightarrow M^j \given M^{[ij]})
    \label{eq:cd}%
    \end{align}
    \vspace{-20pt}
    \begin{enumerate}
      \item For \textbf{discrete variables}:
    \begin{align*}
    \begin{split}
      TE_\tau&(X^i \rightarrow X^j \given X^{[ij]}) = \\ & \displaystyle\sum_{x, x'} p \left(X^j_{t+\tau} = x'^j, X_{t} = x\right) \log \left( \frac{p \left(X^j_{t+\tau} = x'^j \given X_{t} = x \right)}{p \left(X^j_{t+\tau} = x'^j \given X^j_{t} = x^j, X^{[ij]}_{t} = x^{[ij]} \right)} \right)
    \end{split}
    \end{align*}
    \item For \textbf{continuous, linear-Gaussian variables}:
    \begin{align*}
      TE_\tau(X^i \rightarrow X^j \given X^{[ij]}) = \frac{1}{2} \log \left( \frac{\det \Sigma \left(X^j_{t+\tau} \given X^j_t \oplus X^{[ij]}_t \right) }{\det \Sigma \left(X^j_{t+\tau} \given X_t \right)} \right)
    \end{align*}
    \item For \textbf{continuous variables} with an arbitrary distribution, we
    must resort to the nearest-neighbour methods introduced by
    \cite{Kraskov2004}. See reference for details.
  \end{enumerate}
  \label{box:cd}
\end{titledBox}

\subsection{Other measures}

As already mentioned, all the measures reviewed here (besides CD) were inspired
by the $\Phi_{2008}$ measure, which arose from the version of IIT laid out in
Ref.~\cite{Balduzzi2008a}. The most recent version of IIT \cite{Oizumi2014} is
conceptually distinct, and the associated ``$\Phi$-3.0'' is consequently
different to the measures we consider here. The consideration of perturbation
of the system, as well as all of its subsets, in both the past and the future
renders $\Phi$-3.0 considerably more computationally expensive than other
$\Phi$ measures. We do not here attempt to consider the construction of an
analogue of $\Phi$-3.0 for spontaneous information dynamics. Such an
undertaking lies beyond the scope of this paper.

Recently, Tegmark \cite{Tegmark2016} developed a comprehensive taxonomy of all
integrated information measures that can be written as a distance between a
probability distribution pertaining to the whole and one obtained as a product
of probability distributions pertaining to the parts. Tegmark further
identified a shortlist of candidate measures, based on a set of explicit
desiderata. This shortlist overlaps with the measures we consider here, and
also contains other measures which are minor variants. Of Tegmark's shortlisted
measures, $\phi^{\mathrm{M}}$ is equivalent to $\tilde{\Phi}$ under the
system's spontaneous distribution, $\phi^{\mathrm{M}}_{kk'}$ is its
state-resolved version, $\phi^{\mathrm{oak}}$ is transfer entropy (which we
cover here through CD), and $\phi^{\mathrm{npk}}$ is not defined for continuous
variables. The measures $\Phi_G$ and $\psi$ are outside of Tegmark's
classification scheme.

\section{Results}
\label{sec:results}

All of the measures of integrated information that we have described have the
potential to behave in ways which are not obvious \textit{a priori}, and in a
manner difficult to express analytically. While some simulations of $\Phi$,
$\tilde{\Phi}$ and CD on networks have been performed
\cite{Barrett2011,Seth2011}, $\Phi^*$ and $\Phi_G$ have not previously been
computed on models consisting of more than two components, and $\psi$ hasn't
previously been explored at all on systems with continuous variables. In this
section, we study all the measures together on small networks. We compare
the behaviour of the measures, and assess the extent to which each measure is
genuinely capturing dynamical complexity.

To recap, we consider the following 6 measures:
\begin{itemize}[itemsep=-0.3ex]
  \item Whole-minus-sum integrated information, $\Phi$.
  \item Integrated stochastic interaction, $\tilde{\Phi}$.
  \item Decoder-based integrated information, $\Phi^*$.
  \item Geometric integrated information, $\Phi_G$.
  \item Integrated synergy, $\psi$.
  \item Causal density, CD.
\end{itemize}

We use models based on stochastic linear auto-regressive (AR) processes with
Gaussian variables. These constitute appropriate models for testing the
measures of integrated information. They are straightforward to parameterise
and simulate, and are amenable to the formulae presented in Section
\ref{sec:integrationMeasures}. Mathematically, we define an AR process (of
order 1) by the update equation
\begin{equation}
  X_{t+1} = A X_t + \varepsilon_t ,
  \label{eq:ar}%
\end{equation}
where $\varepsilon_t$ is a serially independent random sample from a zero-mean
Gaussian distribution with given covariance $\Sigma(\varepsilon)$, usually
referred to as the \textit{noise} or \textit{error term}. A particular AR
process is completely specified by the coupling matrix or \textit{network} $A$
and the noise covariance matrix $\Sigma(\varepsilon)$. An AR process is stable,
and stationary, if the spectral radius of the coupling matrix is less than 1
\cite{Lutkepohl2005}. (The spectral radius is the largest of the absolute
values of its eigenvalues.) All the example systems we consider are calibrated
to be stable, so the $\Phi$ measures can be computed from their stationary
statistics.

We shall consider how the measures vary with respect to: (i) the strength of
connections, i.e.~the magnitude of non-zero terms in the coupling matrix; (ii)
the topology of the network, i.e~the arrangement of the non-zero terms in the
coupling matrix; (iii) the density of connections, i.e.~the density of non-zero
terms in the coupling matrix; and (iv) the correlation between noise inputs to
different system components, i.e.~the off diagonal terms in
$\Sigma(\varepsilon)$. The strength and density of connections can be thought
of as reflecting, in different ways, the level of integration in the network.
The correlation between noise inputs reflects (inversely) the level of
segregation, in some sense. We also, in each case, compute the control measures
\begin{itemize}[itemsep=-0.3ex]
  \item Time-delayed mutual information (TDMI), $I(X_{t-\tau}, X_t)$; and
  \item Average absolute correlation $\bar\Sigma$, defined as the average
  absolute value of the non-diagonal entries in the system's correlation
  matrix.
\end{itemize}
These simple measures quantify straightforwardly the level of interdependence
between elements of the system, across time and space respectively. TDMI
captures the total information generated as the system transitions from one
time-step to the next, and $\bar\Sigma$ is another basic measure of the level
of integration.

We report the unnormalised measures minimised over even-sized bipartitions --
i.e. bipartitions in which both parts have the same number of components. In
doing this we avoid conflating the effects of the choice of definition of
effective information with those of the choice of partition search (see
Sec.~\ref{sec:mip}). See Discussion (Sec.~\ref{sec:partitions}) for more on
this.

\subsection{Key quantities for computing the integrated information measures}

To compute the integrated information measures, the stationary covariance and
lagged partial covariance matrices are required. By taking the expected value
of $X_t^T X_t$ with Eq.~\eqref{eq:ar} and given that $\varepsilon_t$ is white
noise, uncorrelated in time, one obtains that the stationary covariance matrix
$\Sigma(X)$ is given by the solution to the discrete-time Lyapunov equation,
\begin{align}
  \Sigma (X_t) = A ~ \Sigma (X_t) ~ A^{\mathrm{T}} + \Sigma (\epsilon_t) ~ .
  \label{eq:arcov}%
\end{align}
\noindent This can be easily solved numerically, for example in Matlab via use
of the \texttt{dlyap} command. The lagged covariance can also be calculated
from the parameters of the AR process as
\begin{align}
  \Sigma(X_{t-1}, X_t) = \langle X_t (A X_t + \varepsilon_t)^{\mathrm{T}} \rangle = \Sigma(X_t) A^{\mathrm{T}} ~ ,
\end{align}
\noindent and partial covariances can be obtained by applying
Eq.~\eqref{eq:partialCov}. Finally, we obtain the analogous quantities for the
partitions by the marginalisation properties of the Gaussian distribution.
Given a bipartition $X_t = \{M_t, N_t\}$, we write the covariance and lagged
covariance matrices as
\begin{align}
\begin{split}
  & \Sigma(X_t) = \begin{pmatrix} \Sigma(X_t)_{mm} & \Sigma(X_t)_{mn} \\ \Sigma(X_t)_{nm} & \Sigma(X_t)_{nn} \end{pmatrix} ~ , \\
  & \Sigma(X_{t-1},X_t) = \begin{pmatrix} \Sigma(X_{t-1},X_t)_{mm} & \Sigma(X_{t-1},X_t)_{mn} \\ \Sigma(X_{t-1},X_t)_{nm} & \Sigma(X_{t-1},X_t)_{nn} \end{pmatrix} ~ ,
\end{split}
\end{align}
\noindent and we simply read the partition covariance matrices as
\begin{align}
\begin{split}
  & \Sigma(M_t) = \Sigma(X_t)_{mm} ~ , \\
  & \Sigma(M_{t-1},M_t) = \Sigma(X_{t-1},X_t)_{mm} ~ .
\end{split}
\end{align}

\subsection{Two-node network}
\label{sec:two-node}

\begin{figure}[ht]
  \centering
  \includetikz{tikz/}{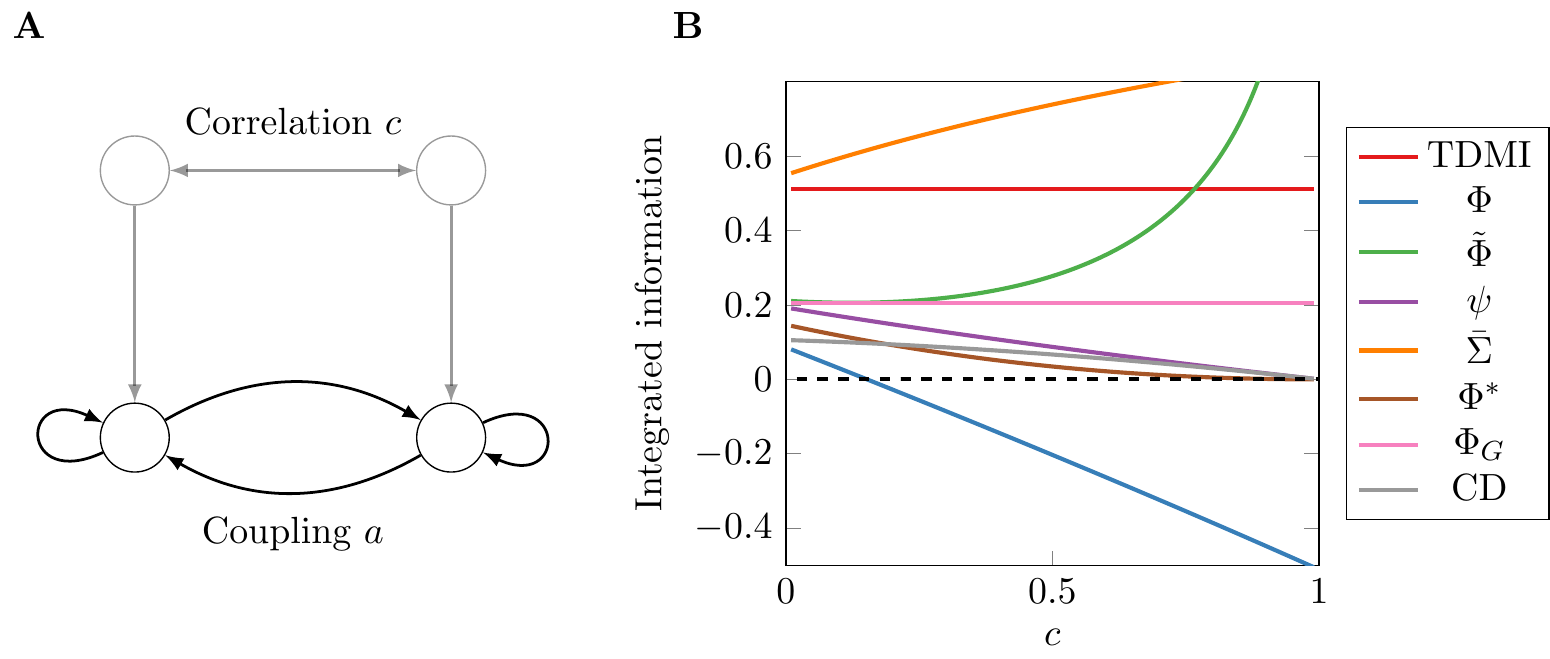}

  \caption{(\textbf{A}) Graphical representation of the two-node AR process
    described in Eq.~\eqref{eq:arNoise}. Two connected nodes with coupling
    strength $a$ receive noise with correlation $c$, which can be thought of as
    coming from a common source. (\textbf{B}) All integrated information measures
    for different noise correlation levels $c$.}

  \label{fig:twoNode1D}
\end{figure}

We begin with the simplest non-trivial AR process,
\begin{subequations}
  \begin{align}
  A ~ & = \begin{pmatrix} a & a \\ a & a \end{pmatrix} ~ , \\
  \Sigma (\epsilon) & = \begin{pmatrix} 1 & c \\ c & 1 \end{pmatrix} ~ .
  \end{align}
  \label{eq:arNoise}%
\end{subequations}
Setting $a = 0.4$ we obtain the same model as depicted in Fig.~3 in
Ref.~\cite{Oizumi2015a}. We simulate the AR process with different levels of
noise correlation $c$ and show results for all the measures in
Fig.~\ref{fig:twoNode1D}. Note that as $c$ approaches 1 the system
becomes degenerate, so some matrix determinants in the formulae become zero
causing some measures to diverge.

Inspection of Figure 2 immediately reveals a wide variability of behaviour
among the measures, in both value and trend, even for this minimally simple
model. Nevertheless, some patterns emerge. Both TDMI and $\Phi_G$ are unaffected by
noise correlation, and both $\tilde{\Phi}$ and $\bar\Sigma$ grow monotonically
with $c$. In fact, $\tilde{\Phi}$ diverges to infinity as $c \rightarrow 1$.
The measures $\psi$, $\Phi^*$, and CD decrease monotonically to 0 when the
effect of the coupling cannot be distinguished from the noise. On the other
hand, $\Phi$ also decreases monotonically but becomes negative for large enough
$c$.

\begin{figure}[ht]
  \begin{adjustwidth}{-0.45in}{0in}
  \centering
  \includetikz{tikz/}{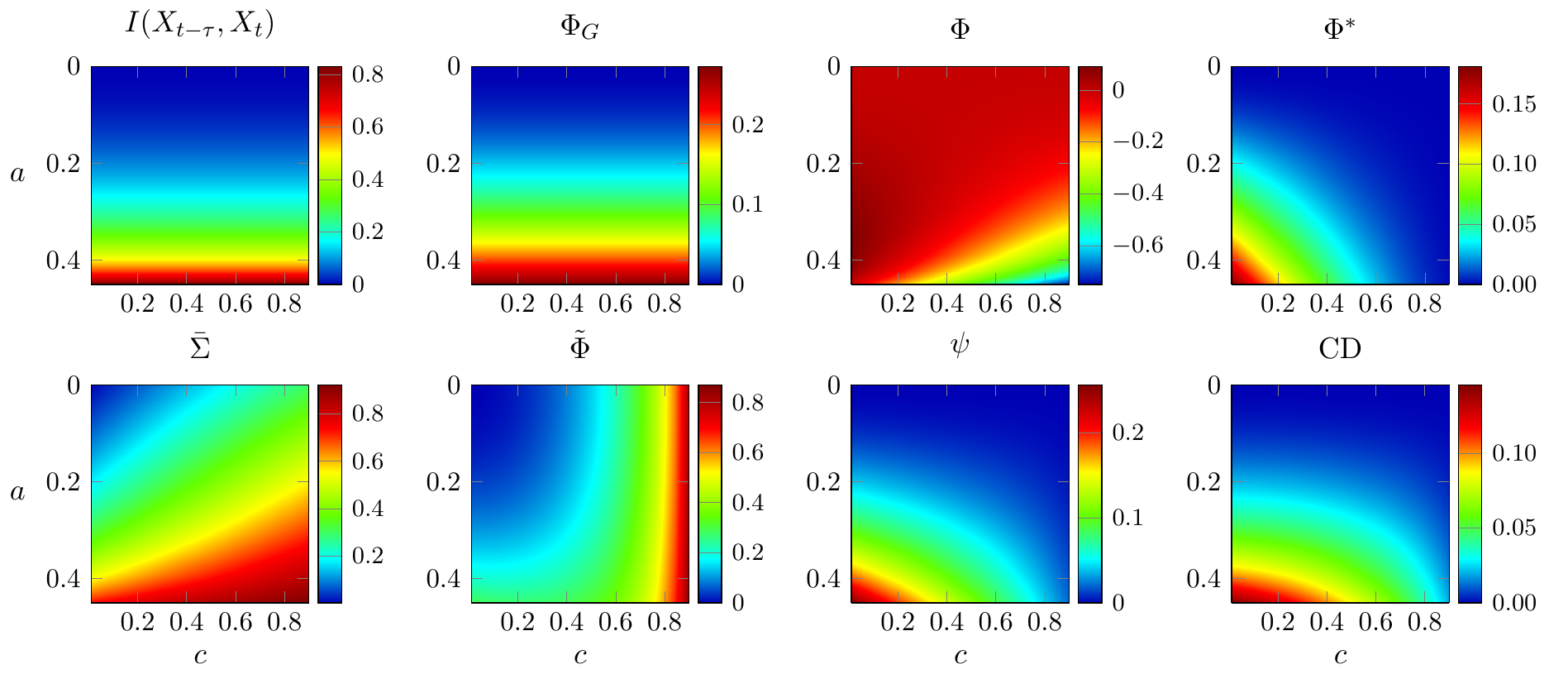}
  \caption{All integrated information measures for the two-node AR process
  described in Eq.~\eqref{eq:arNoise}, for different coupling strengths $a$ and
  noise correlation levels $c$. Vertical axis is inverted for visualisation
  purposes.}
  \label{fig:twoNode2D}
  \end{adjustwidth}
\end{figure}

In Fig.~\ref{fig:twoNode2D} we analyse the same system, but now
varying both noise correlation $c$ and coupling strength $a$. As per the
stability condition presented above, any value of $a \geq 0.5$ makes the
system's spectral radius greater than or equal to 1, so the system becomes
non-stationary and variances diverge. Hence in these plots we evaluate all
measures for values of $a$ below the limit $a=0.5$.

Again, the measures behave very differently. In this case TDMI and $\Phi_G$
remain unaffected by noise correlation, and grow with increasing coupling
strength as expected. In contrast, $\tilde{\Phi}$ and $\bar\Sigma$ increase
with both $a$ and $c$. $\Phi$ decreases with $c$ but shows non-monotonic
behaviour with $a$. Of all the measures, $\psi$, $\Phi^*$, and CD show
desirable properties consistent with capturing conjoined segregation and
integration -- they monotonically decrease with noise correlation and increase
with coupling strength.

\subsection{Eight-node networks}

We now turn to networks with eight nodes, enabling examination of a richer
space of dynamics and topologies.

\begin{figure}[ht]
  \begin{adjustwidth}{-0.45in}{0in}
  \centering
  \includetikz{tikz/}{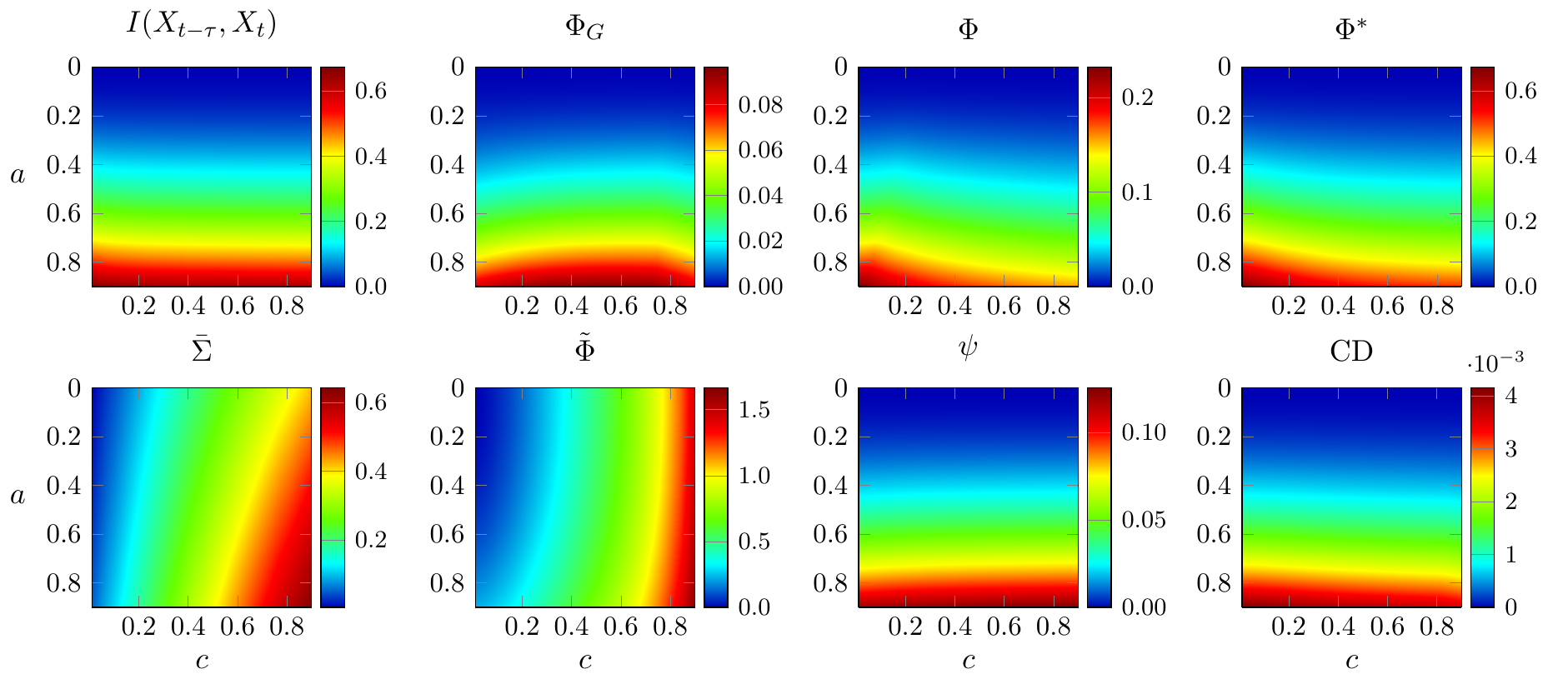}

  \caption{All integrated information measures for the $\Phi$-optimal AR
  process proposed by \cite{Barrett2011}, for different coupling strengths $a$
  and noise correlation levels $c$. Vertical axis is inverted for visualisation
  purposes.}

  \label{fig:optimalNetwork}
  \end{adjustwidth}
\end{figure}

We first analyse a network optimised using a genetic algorithm to yield high
$\Phi$ \cite{Barrett2011}. The noise covariance matrix has ones in the diagonal
and $c$ everywhere else, and now $a$ is a global factor applied to all edges of
the network. The adjacency matrix is scaled such that its spectral radius is
1 when $a = 1$. Similar to the previous section, we evaluate all measures
for multiple values of $a$ and $c$ and show the results in
Fig.~\ref{fig:optimalNetwork}.

Moving to a larger network mostly preserves the features highlighted above.
TDMI is unaffected by $c$; $\tilde{\Phi}$ behaves like $\bar\Sigma$ and diverges for
large $c$; and $\Phi^*$ and CD have the same trend as before, although now the
decrease with $c$ is less pronounced. Interestingly, $\psi$ and $\Phi_G$
increase slightly with $c$, and $\Phi$ does not show the instability and
negative values seen in Fig.~\ref{fig:twoNode2D}.
Overall, in this more complex network the effect of increasing noise
correlation on $\Phi$, $\psi$, $\Phi^*$, and CD is not as pronounced as in
simpler networks, where these measures decrease rapidly towards zero with
increasing $c$.

Thus far we have studied the effect of AR dynamics on integrated
information measures, keeping the topology of the network fixed and changing
only global parameters. We next examine the effect of network
topology, on a set of 6 networks:

\begin{description}[itemsep=-0.3ex]
  \item[A] A fully connected network without self-loops.
  \item[B] The $\Phi$-optimal binary network presented in \cite{Barrett2011}.
  \item[C] The $\Phi$-optimal weighted network presented in \cite{Barrett2011}.
  \item[D] A bidirectional ring network.
  \item[E] A ``small-world'' network, formed by introducing two long-range
  connections to a bidirectional ring network.
  \item[F] An unidirectional ring network.
\end{description}

\begin{figure}[ht]
  \centering
  \includetikz{tikz/}{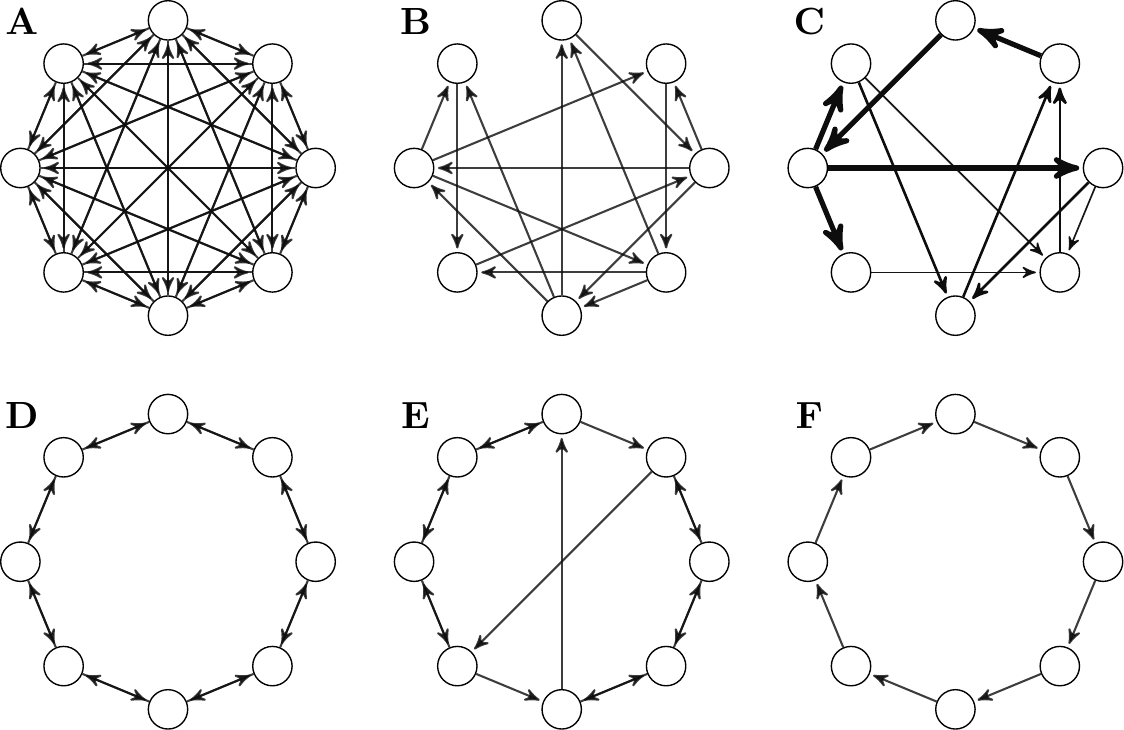}
  \caption{Networks used in the comparative analysis of
    integrated information measures. (\textbf{A}) Fully connected network, (\textbf{B})
    $\Phi$-optimal binary network from \cite{Barrett2011}, (\textbf{C}) $\Phi$-optimal
    weighted network from \cite{Barrett2011}, (\textbf{D}) bidirectional ring network,
  (\textbf{E}) small world network, and (\textbf{F}) unidirectional ring network.}
  \label{fig:netDiagrams}
\end{figure}

\begin{figure}[ht]
\begin{adjustwidth}{-0.45in}{0in}
  \centering
  \includetikz{tikz/}{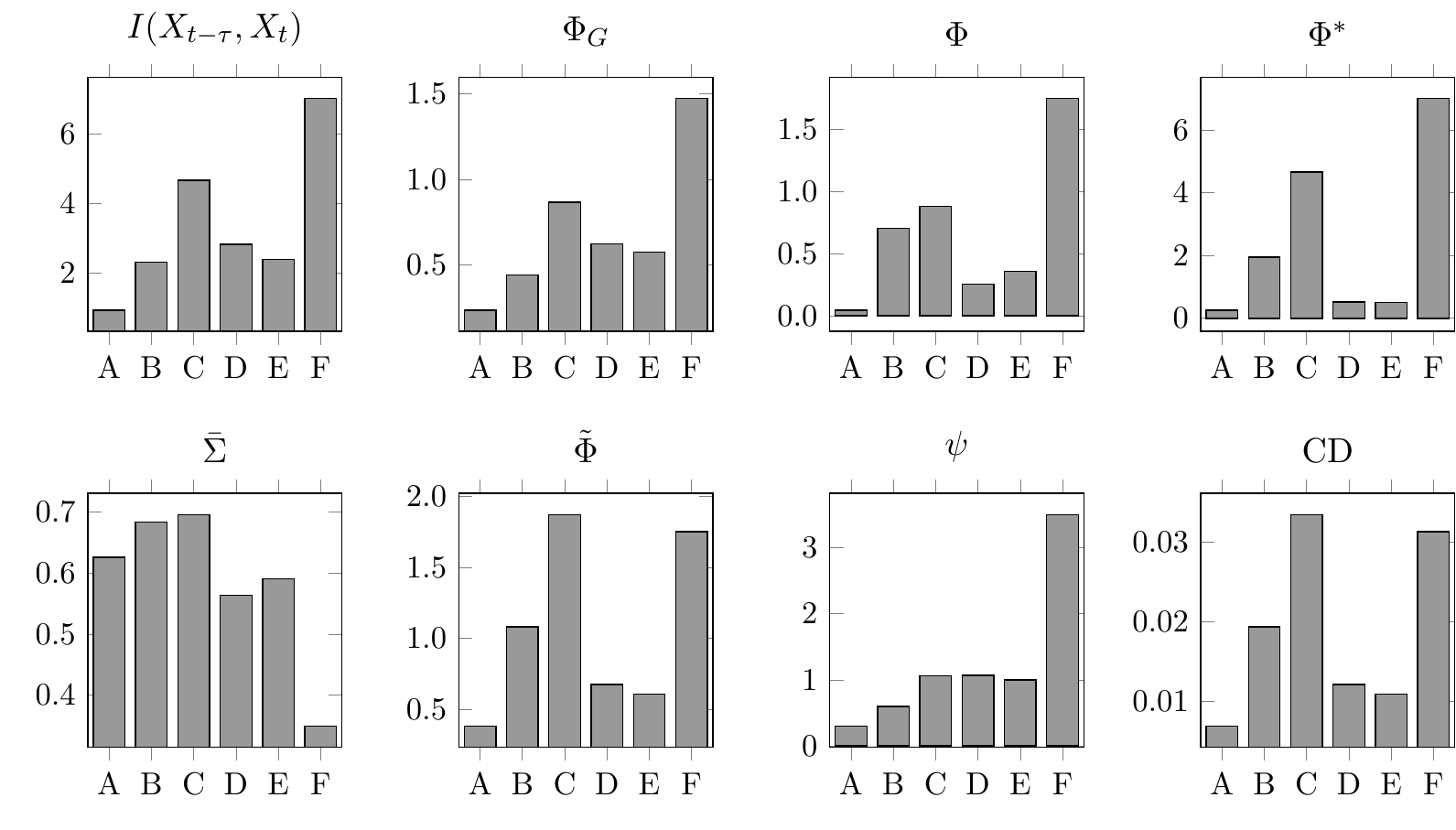}
  \caption{Integrated information measures for all networks in the suite shown in
    Fig.~\ref{fig:netDiagrams}, normalised to spectral radius $0.9$ and under
    the influence of uncorrelated noise. The ring and weighted $\Phi$-optimal
    networks score consistently at the top, while denser networks like the
  fully connected and the binary $\Phi$-optimal networks are usually at the
bottom. Most measures disagree on specific values but agree on the relative
complexity ranking of the networks.}
  \label{fig:netSummary}
\end{adjustwidth}
\end{figure}

In each network the adjacency matrix has been normalised to a spectral radius
of $0.9$. As before, we simulate the system following Eq.~\eqref{eq:ar}, and
here set noise input correlations to zero $(c=0)$ so the noise input covariance
matrix is just the identity matrix. Figure \ref{fig:netDiagrams} shows
connectivity diagrams of the networks for visual comparison, and
Fig.~\ref{fig:netSummary} shows the values of all integrated information
measures evaluated on all networks.

As before, there is substantial variability in the behaviour of all measures,
but some general patterns are apparent. Intriguingly, the unidirectional ring
network is consistently judged by all measures (except for $\tilde{\Phi}$) as
the most complex, followed in most cases by the weighted $\Phi$-optimal
network.\footnote{Note that in Fig.~\ref{fig:netSummary} the $\Phi$-optimal
networks \textbf{B} and \textbf{C} score much less than simpler network
\textbf{F}. This is because all networks have been scaled to a spectral radius
of $0.9$ -- when the networks are normalised to a spectral radius of 0.5, as in
Ref.~\cite{Barrett2011}, then \textbf{B} and \textbf{C} are, as expected, the
networks with highest $\Phi$.
} On the other end of the spectrum, the fully connected network \textbf{A} is
also consistently judged as the least complex network, which is explained
by the large correlation between its nodes as shown by $\bar\Sigma$.

The results here can be summarised by comparing the \textit{relative}
complexity assigned to the networks by each measure -- that is, to what extent
do measures agree on which network is more complex than which. For convenience,
we show the measure-dependent ranking of the network complexity in Table
\ref{tab:netRankings}.

\begin{table}[!ht]
\centering
\caption{Networks ranked according to their value of each integrated
information measure (highest value to the left). We add small-world index
as a dynamics-agnostic measure of network complexity.}
\label{tab:netRankings}
\begin{tabular}{c|c c c c c c}
  \textbf{Measure} & \multicolumn{6}{c}{\textbf{Ranking}} \\
  \toprule
  $I(X_t, X_{t+\tau})$ & F & C & D & E & B & A \\
  $\Phi_G$             & F & C & D & E & B & A \\
  $\Phi$               & F & C & B & E & D & A \\
  $\Phi^*$             & F & C & B & E & D & A \\
  $\bar\Sigma$         & C & B & A & E & D & F \\
  $\tilde{\Phi}$       & C & F & B & D & E & A \\
  $\psi$               & F & C & D & E & B & A \\
  CD                   & C & F & B & D & E & A \\
  \midrule
  SWI                  & C & E & B & A & D & F \\
\end{tabular}
\end{table}

Inspecting this table reveals a remarkable alignment between TDMI, $\Phi_G$,
$\Phi^*$, and $\psi$, especially given how much their behaviour diverges when
varying $a$ and $c$. Although the particular values are different, the measures
largely agree on the ranking of the networks based on their integrated
information. This consistency of ranking is initially encouraging with regard
to empirical application. However, the ranking is not what might be expected
from topological complexity measures from network theory. If we ranked these
networks by e.g. small-world index, we expect networks \textbf{B}, \textbf{C},
and \textbf{E} to be at the top and networks \textbf{A}, \textbf{D}, and
\textbf{F} to be at the bottom -- very different from any of the rankings in
Table \ref{tab:netRankings}.\footnote{The small-world index of a network is
defined as the ratio between its clustering coefficient and its mean minimum
path length, normalised by the expected value of these measures on a random
network of the same density \cite{Humphries2008}. Since the networks we
consider are small and sparse, we use the 4$^{\mathrm{th}}$-order cliques
(instead of triangles, which are 3$^{\mathrm{rd}}$-order cliques) to calculate
the clustering coefficient \cite{Yin2017}.} In fact, the Spearman correlation
between the ranking by small-world index and those by TDMI, $\Phi_G$, $\Phi^*$,
and $\psi$ is around $-0.4$, leading to the counterintuitive conclusion that
more complex networks in fact integrate \textit{less} information. We note that
these rankings are very robust to noise correlation (results not shown) for all
measures except $\Phi$. Indeed, across all simulations in this study the
behaviour of $\Phi$ is erratic, undermining prospects for empirical
application. (This behaviour is even more prevalent if $\Phi$ is optimised over
all bipartitions, as opposed to over even bipartitions.)

\subsection{Random networks}

We next perform a more general analysis of the performance of measures of
integrated information, using Erd\H{o}s-R\'enyi random networks. We consider
Erd\H{o}s-R\'enyi random networks parametrised by two numbers: the edge
density of the network $\rho$ and the noise correlation $c$ (defined as above),
both in the $[0,1)$ interval. To sample a network with a given $\rho$, we
generate a matrix in which each possible edge is present with probability
$\rho$ and then remove self-loops. The stochasticity in the construction of the
Erd\H{o}s-R\'enyi network induces fluctuations on the integrated information
measures, such that for each $(\rho, c)$ we calculate the mean and variance of
each measure.

\begin{figure}[ht]
  \begin{adjustwidth}{-0.45in}{0in}
  \centering
  \includetikz{tikz/}{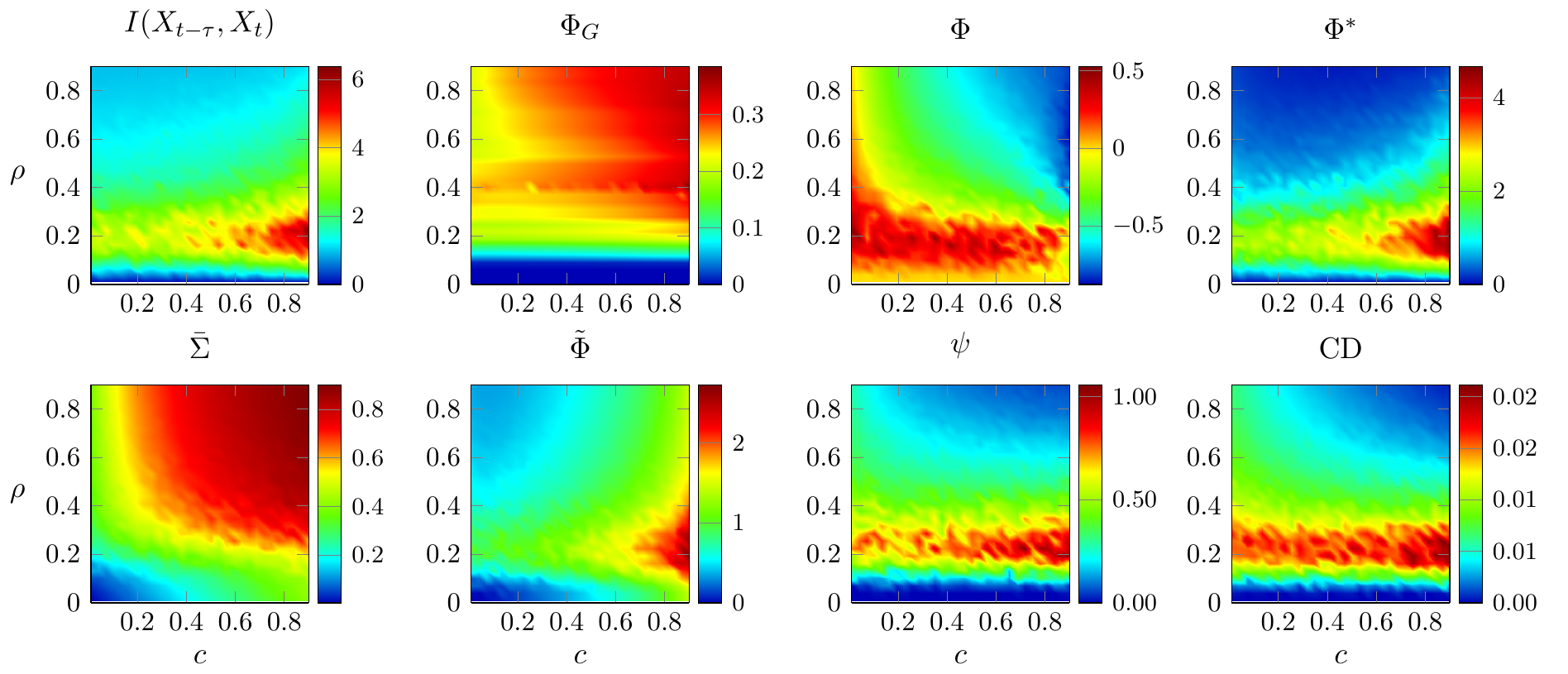}

  \caption{Average integrated information measures for \erdosrenyi random networks
  with given density $\rho$ and noise correlation $c$.}

  \label{fig:erdosRenyi}
  \end{adjustwidth}
\end{figure}

First, we generate 50 networks for each point in the $(\rho,c)$ plane and take
the mean of each integrated information measure evaluated on those 50 networks.
As before, the adjacency matrices are normalised to a spectral radius of $0.9$.
Results are shown in Fig.~\ref{fig:erdosRenyi}. 

$\Phi_G$ increases markedly with $\rho$ and moderately with $c$, $\bar\Sigma$
increases sharply with both and the rest of the measures can be divided in two
groups, with $\Phi$, $\psi$ and CD that decrease with $c$ and TDMI, $\tilde{\Phi}$
and $\Phi^*$ that increase. Notably, all integrated information measures except $\Phi_G$ show a band of
high value at an intermediate value of $\rho$. This demonstrates their
sensitivity to the level of integration. The decrease when $\rho$ is increased
beyond a certain point is due to the weakening of the individual connections in
that case (due to the fixed overall coupling strength, as quantified by
spectral radius).

\begin{figure}[ht]
  \centering
  \includetikz{tikz/}{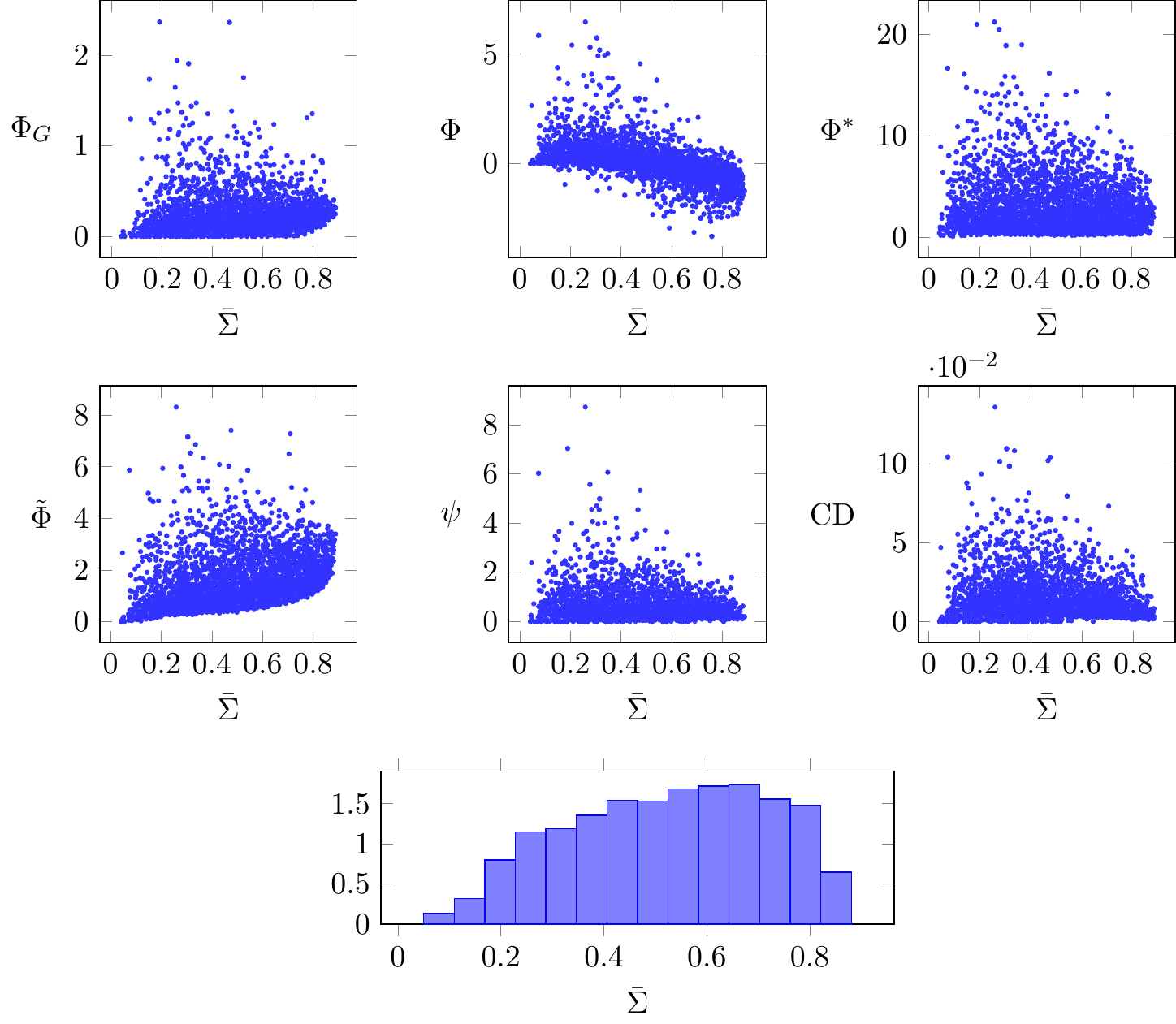}
  \caption{Integrated information measures of random \erdosrenyi networks, plotted against
    the average correlation $\bar\Sigma$ of the same network. (bottom)
  Normalised histogram of $\bar\Sigma$ for all sampled networks.}
  \label{fig:randomNetwork}
\end{figure}

Secondly, in Fig.~\ref{fig:randomNetwork} we plot each
measure against the average correlation of each network, following the
rationale that a good complexity index should peak at an intermediate value of
$\bar\Sigma$ -- i.e. it should reach its maximum value in the middle range of
$\bar\Sigma$. To obtain this figure we sampled a large number of Erd\H{o}s-R\'enyi networks with random
$(\rho, c)$, and evaluated all integrated information measures, as well as their
average correlation $\bar\Sigma$.

Fig.~\ref{fig:randomNetwork} shows that some of the measures have this intermediate peak, in particular: $\Phi^*$, $\psi$, $\Phi_G$, and CD.
Although also showing a modest intermediate peak, $\tilde{\Phi}$ has a stronger
overall positive trend with $\bar\Sigma$, and $\Phi$ an overall negative trend.
These analyses further support $\Phi^*$, $\psi$, $\Phi_G$, and CD
as valid complexity measures, although the relation between them remains
unclear and not always consistent in other scenarios.

One might worry that these peaks could be due to a biased sampling of the
$\bar\Sigma$ axis -- if our sampling scheme were obtaining many more samples
in, say, the $0.2 < \bar\Sigma < 0.4$ range, then the points with high $\Phi$
we see in that range could be explained by the fact that the high-$\Phi$ tails
of the distribution are sampled better in that range than in the rest of the
$\bar\Sigma$ axis. However, the histogram at the bottom of
Fig.~\ref{fig:randomNetwork} shows this is not the case -- on the contrary, the
samples are relatively uniformly spread along the axis. Therefore, the peaks
shown by $\Phi^*$, $\psi$, $\Phi_G$, and CD are not sampling artefacts.

\section{Discussion}

In this study we compared several candidate measures of integrated information
in terms of their theoretical construction, and their behaviour when applied to
the dynamics generated by a range of non-trivial network architectures. We
found that no two measures had precisely the same basic mathematical
properties, see Table \ref{tab:overview}. Empirically, we found a striking
variability in the behaviour among the measures even for simple systems, see
Table \ref{tab:measureSummary} for a summary. Of the measures we have
considered, $\psi$, $\Phi^*$ and CD best capture conjoined segregation and
integration on small networks, when animated with Gaussian linear AR dynamics
(Fig.~\ref{fig:twoNode1D}). These measures decrease with increasing
noise input correlation and increase with increasing coupling strength
(Fig.~\ref{fig:optimalNetwork}). Further, on random networks with fixed overall
coupling strength (as quantified by spectral radius), they achieve their
highest scores when an intermediate number of connections are present
(Fig.~\ref{fig:erdosRenyi}). They also obtain their highest scores when the
average correlation across components takes an intermediate value
(Fig.~\ref{fig:randomNetwork}).

\begin{table}[!ht]
\centering
\caption{Integrated information measures considered and brief summary of our results.}
\label{tab:measureSummary}
\begin{tabular}{c|l|c}
  \textbf{Measure} & \textbf{Summary of results} \\
  \toprule
  $\Phi$         & Erratic behaviour, negative when nodes are strongly correlated. \\
  $\tilde{\Phi}$ & Mostly reflects noise input correlation, not sensitive to changes in coupling. \\
  $\psi$         & Consistent with reflecting both segregation and integration. \\
  $\Phi^*$       & Consistent with reflecting both segregation and integration. \\
  $\Phi_G$       & Mostly reflects changes in coupling, not sensitive to noise input correlation. \\
  CD             & Consistent with reflecting both segregation and integration. \\
\end{tabular}
\end{table}

In terms of network topology, none of the measures strongly reflect complexity
of the network structure in a graph theoretic sense. At fixed overall coupling
strength, a simple ring structure (Fig.~\ref{fig:netDiagrams}) leads in most
cases to the highest scores. Among the other measures: $\tilde{\Phi}$ is
largely determined by the level of correlation amongst the noise inputs, and is
not very sensitive to changes in coupling strength; $\Phi_G$ depends mainly on
the overall coupling strength, and is not very sensitive to changes in noise
input correlation; and $\Phi$ generally behaves erratically.

Considered together, our results motivate the continued development of $\psi$,
$\Phi^*$ and CD as theoretically sound and empirically adequate measures of
integrated information.

\subsection{Partition selection}
\label{sec:partitions}

Integrated information is typically defined as the \textit{effective
information} beyond the \textit{minimum information partition}
\cite{Balduzzi2008a,Tononi2003}. However, when a particular measure of
integrated information has been first introduced, it is often with a new
operationalisation of both effective information and the minimum information
partition. In this paper we have restricted attention to comparing different
choices of measure of \textit{effective information}, while keeping the same
partition selection scheme across all measures. Specifically, we restricted the
partition search to even-sized bipartitions, which has the advantage of
obviating the need for introducing a normalisation factor when comparing
bipartitions with different sizes, see Section \ref{sec:mip}. For uneven
partitions, normalisation factors are required to compensate for the fact that
there is less capacity for information sharing as compared to even partitions.
However, such factors are known to introduce instabilities, both under
continuous parameter changes, and in terms of numerical errors
\cite{Barrett2011}. Further research is needed to compare different approaches
to defining the minimum information partition, or finding an approximation to
it in reasonable computation time \cite{Toker2016}.

In terms of computation time, performing the most thorough search, through all
partitions, as in the early formulation of $\Phi$ by Balduzzi and Tononi
\cite{Balduzzi2008a} requires time $\mathcal{O}(n^n)$\footnote{More precisely,
as the Bell number $B_n$.}. Restricting attention to bipartitions reduces this
to $\mathcal{O}(2^n)$, whilst restricting to even bipartitions reduces this
further to $\mathcal{O}(n^2)$. These observations highlight a trade-off between
computation time and comprehensive consideration of possible partitions. Future
comparisons of integrated information measures may benefit from more advanced
methods for searching among a restricted set of partitions to obtain a good
approximation to the minimum information partition. For example, Toker and
Sommer use graph modularity, stochastic block models or spectral clustering as
informed heuristics to suggest a small number of partitions likely to be close
to the MIP, and then take the minimum over those. With these approximations
they are able to calculate the MIP of networks with hundreds of nodes
\cite{Toker2016,Toker2017}. Alternatively, Hidaka and Oizumi make use of the
submodularity of mutual information to perform efficient optimisation and find
the bipartition across which there is the least instantaneous mutual
information of the system \cite{Hidaka2017}. Presently, however, their method
is valid only for instantaneous mutual information and is therefore not
applicable to finding the bipartition that minimises any form of normalised
effective information as described in Section \ref{sec:mip}.

Further, each measure carries special considerations regarding partition
search. For example, for $\psi$, taking the minimum across all partitions is
equivalent to taking it across bipartitions only, thanks to the properties of
$I_\cap$ \cite{Williams2010,Barrett2014,Rosas2016}. Arsiwalla and Verschure
\cite{Arsiwalla2013} used $\tilde{\Phi}$ and suggested always using the atomic
partition on the basis that it is fast, well-defined, and for $\tilde{\Phi}$
specifically it can be proven to be the partition of \textit{maximum}
information; and thus it provides a quickly computable upper bound for the
measure.

\subsection{Continuous variables and the linear Gaussian assumption}

We have compared the various integrated information measures only on systems
whose states are given by continuous variables with a Gaussian distribution.
This is motivated by measurement variables being best characterised as
continuous in many domains of potential application. Future research should
continue the comparison of these measures on a test-bed of systems with
discrete variables. Moreover, non-Gaussian continuous systems should also be
considered because the Gaussian approximation is not always a good fit to real
data. For example, the spiking activity of populations of neurons typically
exhibit exponentially distributed dynamics \cite{Dayan2001}. Systems with
discrete variables are in principle straightforward to deal with, since
calculating probabilities (following the most brute-force approach) amounts
simply to counting occurrences of states. General continuous systems, however,
are less straightforward. Estimating generic probability densities in a
continuous domain is challenging, and calculating information-theoretic
quantities on these is difficult \cite{Kraskov2004,Wang2009a}. The AR systems
we have studied here are a rare exception, in the sense that their probability
density can be calculated and all relevant information-theoretic quantities
have an analytical expression. Nevertheless, the Gaussian assumption is common
in biology, and knowing now how these measures behave on these Gaussian systems
will inform further development of these measures, and motivate their
application more broadly.

\subsection{Empirical as opposed to maximum entropy distribution}
\label{sec:maxent_discussion}

We have considered versions of each measure that quantify information with
respect to the empirical, or spontaneous, stationary distribution for the state
of the system. This constitutes a significant divergence from the supposedly
fundamental measures of intrinsic integrated information of IIT versions 2 and
3 \cite{Balduzzi2008a,Oizumi2014}. Those measures are based on information
gained about a hypothetical past moment in which the system was equally likely
to be in any one of its possible states (the `maximum entropy' distribution).
However, as pointed out previously \cite{Barrett2011}, it is not possible to
extend those measures, developed for discrete Markovian systems, to continuous
systems. This is because there is no uniquely defined maximum entropy
distribution for a continuous random variable (unless it has hard-bounds,
i.e.~a closed and bounded set of states). Hence, quantification of information
with respect to the empirical distribution is the pragmatic choice for
construction of an integrated information measure applicable to continuous
time-series data.

The consideration of information with respect to the empirical, as opposed to
maximum entropy, distribution does however have an effect on the concept
underlying the measure of integrated information -- it results in a measure not
of mechanism, but of dynamics \cite{Barrett2013}. That is, what is measured is
not information about what the possible mechanistic causes of the current state
\textit{could be}, but rather what the likely preceding states \textit{actually
are}, on average, statistically; see \cite{Barrett2011} for further discussion.
Given the diversity of behaviour of the various integrated information measures
considered here even on small networks with linear dynamics, one must remain
cautious about considering them as generalisations or approximations of the
proposed `fundamental' $\Phi$ measures of IIT versions 2 or 3
\cite{Balduzzi2008a,Oizumi2014}.

A remaining important challenge, in many practical scenarios, is the identification of
stationary epochs. For a relatively long data segment, it can be unrealistic to
assume that all the statistics are constant throughout. For shorter data
segments, one can not be confident that the system has explored all the states
that it potentially would have, given enough time.  

\section{Final remarks}

The further development, and empirical application of Integrated Information
Theory requires a satisfactory informational measure of dynamical complexity.
During the last few years several measures have been proposed, but their
behaviour in any but the simplest cases has not been extensively characterised
or compared. In this study, we have reviewed several candidate measures of
integrated information, and provided a comparative analysis on simulated data,
generated by simple Gaussian dynamics applied to a range of network topologies.

Assessing the degree of dynamical complexity, integrated information, or
co-existing integration and segregation exhibited by a system remains an
important outstanding challenge. Progress meeting this challenge will have
implications not only for theories of consciousness, such as Integrated
Information Theory, but more generally in situations where relations between
local and global dynamics are of interest. The review presented here identifies
promising theoretical approaches for designing adequate measures of integrated
information. Further, our simulations demonstrate the need for empirical
investigation of such measures, since measures that share similar theoretical
properties can behave in substantially different ways, even on simple systems.

\section*{Acknowledgements}

The authors would like to thank Michael Schartner for advice. ABB is funded by
EPSRC grant EP/L005131/1. ABB and AKS are grateful to the Dr. Mortimer and
Theresa Sackler Foundation, which supports the Sackler Centre for Consciousness
Science. AKS is additionally grateful to the CIFAR Azrieli programme on Mind,
Brain, and Consciousness.

\appendix

\section*{Appendix}

\section{Derivation and concavity proof of $I^*$}
\label{ap:phiStar}

\subsection{Derivation of $I^*$ in Gaussian systems}

Here we provide a closed-form expression for the mismatched decoding
information in a Gaussian dynamical system. See Section \ref{sec:phiStar} for
more information. For clarity, we omit the $X, \tau, \mathcal{P}$ arguments of
$\tilde{I}$ and write it as a function of $\beta$ only. The formula for
$\tilde{I}(\beta)$ for a stationary continuous random process is
\begin{equation}
  \tilde{I}(\beta)=-\int \dd x\, p(x) \log \int \dd \tix\, p(\tix)q(x|\tix)^\beta + \int \dd \tix \int \dd x\, p(x,\tix)\log q(x|\tix)^\beta\,,
  \label{eq:integrals}%
\end{equation}
where $p(x)$ is the distribution for $X_t$, $p(x,\tix)$ is the joint
distribution for $(X_t,X_{t-\tau})$, and $q(x|\tix)$ is the conditional
distribution for $X_t$ given $X_{t-\tau}$ under the partitioning in question.
The function $\tilde{I}(\beta)$ also depends on $X_t$, $\tau$ and
$\mathcal{P}$, but for the sake of clarity we omit all arguments except for
$\beta$, which is the parameter of interest here. When $X_t$ is Gaussian with
covariance matrix $\Sigma_X$ (and mean 0 without loss of generality), we have
\begin{equation}
  p(x)=(2\pi)^{-n/2} | \Sigma_X |^{-1/2} \mathrm{exp}\left[ -\frac{1}{2} \psi\left(x,\Sigma^{-1}_X\right) \right]\,,
  \label{eq:px}%
\end{equation}
where we define
\begin{equation}
  \psi(x,M)=:x^{\mathrm{T}}Mx
\end{equation}
for a vector $x$ and a matrix $M$. Further
\begin{eqnarray}
  q(x|\tix)&=&(2\pi)^{-n/2} | \Pi_{X|\tiX} |^{-1/2} \mathrm{exp}\left[ -\frac{1}{2} \psi\left(x-\Pi_{X\tiX}\Pi_X^{-1}\tix,\Pi^{-1}_{X|\tiX}\right) \right]\,,
  \label{eq:qx}%
\end{eqnarray}
where $\Pi_X$ is the block diagonal covariance matrix for $X_t$ under the
partition, $\Pi_{X\tiX}=:\Sigma_q(X_t,X_{t-\tau})=\Pi_{\tiX X}^{\mathrm{T}}$ is
the block diagonal auto-covariance matrix associated with the partition, and
$\Pi_{X|\tiX}$ is the partial covariance
\begin{equation}
  \Pi_{X|\tiX}=\Pi_X-\Pi_{X\tiX}\Pi_X^{-1}\Pi_{\tiX X}\,.
\end{equation}
We start with the integral
\begin{equation}
  \int \dd \tix\, p(\tix)q(x|\tix)^\beta  = (2\pi)^{-n\beta/2} | \Pi_{X|\tiX} |^{-\beta/2}(2\pi)^{-n/2} \left|\Sigma_X\right|^{-1/2} \int \dd\tix\,\mathrm{exp} (\mathcal{E})\,,
  \label{eq:calE}%
\end{equation}
where
\begin{equation}
  \mathcal{E}=\frac{1}{2}\tix^\trans\Sigma_X^{-1}\tix-\frac{\beta}{2}\tix^\trans\Pi_X^{-1} \Pi_{\tiX X} \Pi_{X|\tiX}^{-1} \Pi_{X \tiX} \Pi_X^{-1} \tix + \beta x^\trans \Pi_{X|\tiX}^{-1} \Pi_{X \tiX} \Pi_X^{-1} \tix -\frac{\beta}{2} x^\trans \Pi_{X|\tiX}^{-1} x\,.
\end{equation}
If we write
\begin{equation}
  \mathcal{E}=-\frac{1}{2}(\tix-Bx)^\trans Q (\tix-Bx)-\frac{1}{2}x^\trans R_1 x\,,
\end{equation}
then
\begin{subequations}
\begin{align}
Q&= \Sigma_X^{-1}+\beta \Pi_X^{-1} \Pi_{\tiX X} \Pi_{X|\tiX}^{-1} \Pi_{X \tiX} \Pi_X^{-1}\,,\\
B^\trans&= \beta \Pi_{X|\tiX}^{-1} \Pi_{X \tiX} \Pi_X^{-1}Q^{-1}\,,\\
R_1&= \beta \Pi_{X|\tiX}^{-1} -\beta^2 \Pi_{X|\tiX}^{-1} \Pi_{X \tiX} \Pi_X^{-1}Q^{-1}\Pi_X^{-1} \Pi_{\tiX X} \Pi_{X|\tiX}^{-1}\,,
\end{align}
\end{subequations}
so
\begin{eqnarray}
\begin{split}
\int \dd\tix\,\mathrm{exp} (\mathcal{E})&= \mathrm{exp}\left( -\frac{1}{2} x^\trans R_1 x \right) \int \dd y\, \mathrm{exp}\left(-\frac{1}{2} y^\trans Q y \right)\\
&= \mathrm{exp}\left( -\frac{1}{2} x^\trans R_1 x \right) (2\pi)^{n/2}|Q|^{-1/2}\,.
\end{split}
\end{eqnarray}
Hence, using \eqref{eq:px} and \eqref{eq:calE} we obtain the first term in \eqref{eq:integrals}:
\begin{multline}
  -\int \dd x\, p(x) \log \int \dd \tix\, p(\tix)q(x|\tix)^\beta = \frac{n\beta}{2}\log 2\pi \\
  + \frac{1}{2} \log \left( |Q| \cdot |\Sigma_X| \cdot | \Pi_{X|\tiX} |^\beta \right)+\frac{1}{2}\tr (\Sigma_XR_1)\,.
  \label{eq:firstterm}%
\end{multline}
\\

\noindent Now, moving on to the second term in  \eqref{eq:integrals},
\begin{equation}
  \int \dd \tix \int \dd x\, p(x,\tix)\log q(x|\tix)^\beta = -\frac{\beta n}{2} \log 2\pi -\frac{\beta}{2} \log  | \Pi_{X|\tiX} | -\frac{\beta}{2} I_1\,,
\end{equation}
where
\begin{align}
I_1 =& \int \dd \tix \int \dd x\, \, p(x,\tix) \,\psi\left(x-\Pi_{X\tiX}\Pi_X^{-1}\tix,\Pi^{-1}_{X|\tiX}\right) \nonumber\\
=& \int \dd x \, p(x)\,  \psi\left(x, \Pi^{-1}_{X|\tiX}\right ) + \int \dd \tix \, p(\tix) \,\psi\left( \tix, \Pi_X^{-1}\Pi_{\tiX X} \Pi^{-1}_{X|\tiX}\Pi_{X\tiX}\Pi_X^{-1}\right) \nonumber \nonumber\\
 & -2\int \dd \tix \int \dd x\,p(x,\tix)\,x^\trans \Pi^{-1}_{X|\tiX}\Pi_{X\tiX}\Pi_X^{-1}\tix \nonumber\\
=& \tr\left(  \Pi^{-1}_{X|\tiX}\Sigma_X\right)+\tr\left( \Pi_X^{-1}\Pi_{\tiX X} \Pi^{-1}_{X|\tiX}\Pi_{X\tiX}\Pi_X^{-1}\Sigma_X\right) -2\, \tr\left(\Pi^{-1}_{X|\tiX}\Pi_{X\tiX}\Pi_X^{-1}\Sigma_{\tiX X}\right)\,,
\end{align}
where $\Sigma_{\tiX X}=:\Sigma(X_{t-\tau},X_t)$. Thus the second term in
\eqref{eq:integrals} is given by
\begin{multline}
  \int \dd \tix \int \dd x\, p(x,\tix)\log q(x|\tix)^\beta =-\frac{\beta n}{2} \log 2\pi -\frac{\beta}{2} \log  | \Pi_{X|\tiX} | \\ 
  + \frac{1}{2}\tr ( \Sigma_X R_2)+\beta \tr \left(\Pi^{-1}_{X|\tiX}\Pi_{X\tiX}\Pi_X^{-1}\Sigma_{\tiX X}\right)\,,
  \label{eq:secondterm}%
\end{multline}
where
\begin{equation}
  R_2=-\beta \Pi^{-1}_{X|\tiX} -\beta \Pi_X^{-1}\Pi_{\tiX X} \Pi^{-1}_{X|\tiX}\Pi_{X\tiX}\Pi_X^{-1}\,.
\end{equation}
Finally, putting all the terms \eqref{eq:firstterm}, \eqref{eq:secondterm}
together, we obtain
\begin{equation}
  \tilde{I}(\beta)=\frac{1}{2} \log \left( |Q| \cdot |\Sigma_X| \right)+\frac{1}{2}\tr ( \Sigma_X R)+\beta \,\tr \left(\Pi^{-1}_{X|\tiX}\Pi_{X\tiX}\Pi_X^{-1}\Sigma_{\tiX X}\right)\,,
  \label{eq:Ibeta}%
\end{equation}
where
\begin{eqnarray}
Q&=&\Sigma_X^{-1}+\beta \Pi_X^{-1} \Pi_{\tiX X} \Pi_{X|\tiX}^{-1} \Pi_{X \tiX} \Pi_X^{-1}\,,\\
R&=&  -\beta \Pi_X^{-1}\Pi_{\tiX X} \Pi^{-1}_{X|\tiX}\Pi_{X\tiX}\Pi_X^{-1}-\beta^2 \Pi_{X|\tiX}^{-1} \Pi_{X \tiX} \Pi_X^{-1}Q^{-1}\Pi_X^{-1} \Pi_{\tiX X} \Pi_{X|\tiX}^{-1}\,.
\end{eqnarray}
We note that this formula for $\tilde{I}(\beta)$ has been verified with
numerical methods, and it is not the same as the formula reported by Oizumi et
al. \cite{Oizumi2015a}.

\subsection{$\tilde{I}(\beta)$ is concave in $\beta$ in Gaussian systems}

Throughout this proof we will rely multiple times on the the book \emph{Convex
Optimization} by Boyd and Vandenberghe \cite{Boyd2004}. Our aim is to show that
$\tilde{I}(\beta)$ is concave in $\beta$,\footnote{We follow Boyd and
Vandenberghe's notation: a function $f$ is said to be convex, convex downwards
or concave upwards if $f(ax + by) \leq a f(x) + b f(y)$.} which means it has a
unique maximum and can be treated with standard convex optimisation tools.

We start with the second term in Eq.~\eqref{eq:integrals},
\begin{equation}
\int \dd \tix \int \dd x\, p(x,\tix)\log q(x|\tix)^\beta = \beta \int \dd \tix \int \dd x\, p(x,\tix)\log q(x|\tix) \,,
\end{equation}
\noindent which is linear in $\beta$. Moving to the first term, using
Eq.~\eqref{eq:qx} it can be rewritten as
\begin{multline*}
-\int \dd x\, p(x) \log \left[\int \dd \tix\, p(\tix) q(x|\tix)^{\beta}\right]
= -\int \dd x\, p(x) \left[-\frac{n \beta}{2} \log 2 \pi -
\frac{\beta}{2}\log|\Pi_{X|\tiX}|\right] \\ -\int \dd x\, p(x) \log
\left[p(\tix) \exp\left(-\beta f(x, \tix) \right) \dd \tix \right] ~ .
\end{multline*}
We see that the only nonlinear term in $\tilde{I}(\beta)$ is
\begin{equation}
- \int \dd x\, p(x) \log \left[ \int \dd \tix\, p(\tix) \exp(- \beta f(x,\tix) \right] ~ ,
\label{eq:intlogint}%
\end{equation}
\noindent where
\begin{equation}
f(x, \tix) = \frac{1}{2} \psi\left(x-\Pi_{X \tiX} \Pi_X^{-1} \tix, \Pi_{X|\tiX}^{-1}\right) ~ .
\end{equation}
Now we draw from two lemmas presented in \cite{Boyd2004}:
\begin{itemize}
  \item An affine function preserves concavity, in the sense that a linear
  combination of convex (concave) functions is also convex (concave).
  \item A non-negative weighted sum preserves concavity. Since $p(x) > 0$ the
  outer integral in Eq.~\eqref{eq:intlogint} preserves concavity,
\end{itemize}
With these two remarks, we know that to prove the concavity of
$\tilde{I}(\beta)$ we just need to prove the concavity of
\begin{equation}
- \log \left[ \int \dd \tix\, p(\tix) \exp \left( - \beta f(x, \tix) \right) \right] ~ .
\end{equation}
This is known as a \emph{log-sum-exp} function, which as per Sec.~3.1.5 of
\cite{Boyd2004} is convex in $\beta$. Finally, the minus sign in the last
equation flips the convexity and we conclude that $\tilde{I}(\beta)$ is concave
in $\beta$.

\section{Bounds on causal density}
\label{ap:cd}

We now prove that causal density is upper-bounded by time-delayed mutual
information, satisfying what other authors have considered a fundamental
requirement for a measure of integrated information \cite{Oizumi2015}. As
before, we omit the arguments to CD for clarity. We begin by writing down CD in
terms of mutual information:
\begin{align}
\begin{split}
  \mathrm{CD} & = \frac{1}{n(n-1)} \sum_{i \neq j} \mathrm{TE}_{\tau}(X^i \rightarrow X^j | X^{[ij]}) \\
              & = \frac{1}{n(n-1)} \sum_{i \neq j} I(X_t^i ; X_{t+\tau}^j | X_t^{[i]}) ~ ,
\end{split}
\end{align}
\noindent where as before $X_t^{[i]}$ represents the set of all variables in
$X_t$ except $X_t^i$. We will use the chain rule of mutual information
\cite{Cover2006},
\begin{align}
  I(X; Y,Z) = I(X;Z) + I(X;Y|Z) ~ .
  \label{eq:chainRuleMI}%
\end{align}
Using this chain rule and the non-negativity of mutual information we can state
that $I(X_t^i ; X_{t+\tau}^j | X_t^{[i]}) \leq I(X_t; X_{t+\tau}^j)$, and
therefore
\begin{align}
  \mathrm{CD} \leq \frac{1}{n(n-1)} \sum_{i \neq j} I(X_t; X_{t+\tau}^i) ~ .
\end{align}
Also by the same chain rule, it is easy to see that $I(X_t; X_{t+\tau}^i) \leq
I(X_t; X_{t+\tau})$. Then
\begin{align}
  \mathrm{CD} \leq \frac{1}{n(n-1)} \sum_{i\neq j} I(X_t; X_{t+\tau}) ~ .
\end{align}
Given that the sum runs across all $n(n-1)$ pairs, we arrive at our result
\begin{align}
  \mathrm{CD} \leq I(X_t; X_{t+\tau}) ~ .
\end{align}

\section{Properties of integrated information measures}
\label{ap:proofs}

We prove the properties of in Table~\ref{tab:overview}. We will make use of the
properties of mutual information introduced in Sec.~\ref{sec:preliminaries},
repeated here for convenience:

\begin{enumerate}[label=MI-\arabic*]\itemsep0pt
  \item $I(X; Y) = I(Y; X)$, \label{mi:symmetry}
  \item $I(X; Y) \geq 0$, \label{mi:positive}
  \item $I(f(X); g(Y)) = I(X; Y)$ for any injective functions $f, g$, \label{mi:function}
\end{enumerate}

\subsection*{Whole-minus-sum integrated information $\Phi$}

\begin{description}
  \item[Time-symmetric] Follows from \eqref{mi:symmetry}.
  \item[Non-negative] Proof by example. If $X_t^i = X_t^j$, we have $\Phi = (1-N) I(X_t^i; X_{t-\tau}^i) \leq 0$.
  \item[Rescaling-invariant] Follows from \eqref{mi:function} when Balduzzi and Tononi's \cite{Balduzzi2008a} normalisation factor is not used.
  \item[Bounded by TDMI] Follows from \eqref{mi:positive}.
\end{description}

\subsection*{Integrated stochastic interaction $\tilde{\Phi}$}

\begin{description}
  \item[Time-symmetric] Follows from $H(X_t | H_{t-\tau}) = H(X_{t-\tau} | H_t)$, which
  can be proved starting from the system temporal joint entropy
  \begin{align*}
    H(X_t, X_{t-\tau}) & = H(X_t | X_{t-\tau}) + H(X_{t-\tau}) \\
    & = H(X_{t-\tau}, X_t) = H(X_{t-\tau} | X_t) + H(X_t) ~ ,
  \end{align*}
  And using the fact that by the ergodic property $H(X_t) = H(X_{t-\tau})$. The same
  logic applies to all parts of the system.
  \item[Non-negative] Follows from the fact that $\tilde{\Phi}$ is an M-projection (see Ref.~\cite{Oizumi2015}).
  \item[Rescaling-invariant] Follows from the non-invariance of differential entropy \cite{Cover2006} (regardless of whether a normalisation factor is used).
  \item[Bounded by TDMI] Proof by counterexample. In the AR process of Sec.~\ref{sec:two-node} $\tilde{\Phi} \rightarrow \infty$ as $c \rightarrow 1$, although TDMI remains finite.
\end{description}

\subsection*{Integrated synergy $\psi$}

\begin{description}
  \item[Time-symmetric] Proof by counterexample. For the AR system with
  \begin{align*}
    A = \begin{pmatrix} a & a \\ 0 & 0 \end{pmatrix} \qquad,  \qquad
    \Sigma(\varepsilon) = \begin{pmatrix} 1 & 0 \\ 0 & 1 \end{pmatrix}
  \end{align*}
  We have $\psi = \frac{1}{2}\log\left(1 + a^2\right)$ while for the time-reversed
  process $\psi = \frac{1}{2}\log\left(1 + a^4\right)$. Note that this proof applies only
  to the MMI-PID used in this paper and presented in \cite{Barrett2014}.
  \item[Non-negative] Follows from $I_\cup(X, Y; Z) < I(\{X,Y\}; Z)$ \cite{Williams2010}.
  \item[Rescaling-invariant] Follows from \eqref{mi:function} and the
  fact that $I_{\cap}$ is also invariant (see property (\textbf{Eq}) in
  Section 5 of \cite{Griffith2014}).
  \item[Bounded by TDMI] Follows from the non-negativity of $I_\cup$ \cite{Williams2010}.
\end{description}

\subsection*{Decoder-based integrated information $\Phi^*$}

\begin{description}
%
%
  \item[Non-negative] Follows from $I^*[X; \tau, \mathcal{P}] \leq I(X_t;
  X_{t-\tau})$, proven in \cite{Merhav1994}.
  \item[Rescaling-invariant] Assume the measure is computed on a time series
  of rescaled data $X_t^r = X_tA$, where $A$ is a diagonal matrix with positive
  real numbers. Then its covariance is related to the covariance of the original
  time series as $\Sigma_X^r = \mathbb{E}\left[{X_t^r}^\trans X_t^r\right] =
  \mathbb{E}\left[A^\trans X_t^\trans X_t A\right] = A^2 \Sigma_X$. We can analogously
  calculate $\Pi_X, \Pi_{X\tiX}, \Pi_{X|\tiX}$ and easily verify that all $A$'s
  cancel out, proving the invariance.
  \item[Bounded by TDMI] Follows from $I^*[X; \tau, \mathcal{P}] \geq 0$, proven in \cite{Merhav1994}.
\end{description}

\subsection*{Geometric integrated information \phiGeometric}

\begin{description}
  \item[Time-symmetric] Follows from the symmetry in the constraints that
  define the manifold of restricted models $Q$ \cite{Oizumi2015}.
  \item[Non-negative] Follows from the fact that \phiGeometric is an M-projection \cite{Oizumi2015}.
  \item[Rescaling-invariant] Given a Gaussian distribution $p$ with
  covariance $\Sigma_p$, its M-projection in $Q$ is another Gaussian with
  covariance $\Sigma_q$. Given a new distribution $p'$ formed by rescaling
  some of the variables in $p$, the M-projection of $p'$ is
  a Gaussian with covariance $A^2\Sigma_q$ with $A$ a diagonal positive matrix
  (see above),
  which satisfies $D_{KL}(p\|q) = D_{KL}(p'\|q')$ and therefore \phiGeometric
  is invariant to rescaling.

  \item[Bounded by TDMI] TDMI can be defined as the M-projection of the full
  model $p$ to a manifold of restricted models $Q^{MI} = \{q ~ \colon q(X_t,
  X_{t-\tau}) = q(X_t) q(X_{t-\tau})\}$ \cite{Oizumi2015}. The bound
  $\phiGeometric \leq I(X_t; X_{t-\tau})$ follows from the fact that $Q^{MI}
  \subset Q$.

\end{description}

\subsection*{Causal density}

\begin{description}
  \item[Time-symmetric] Follows from the non-symmetry of transfer entropy \cite{Wibral2014}.
  \item[Non-negative] Re-writing CD as a sum of conditional MI terms, follows from \eqref{mi:positive}.
  \item[Rescaling-invariant] Follows from \eqref{mi:function}.
  \item[Bounded by TDMI] Proven in Appendix \ref{ap:cd}.
\end{description}

\bibliographystyle{habbrv}
\bibliography{/home/pmediano/Downloads/Mendeley/library.bib}

\end{document}